\documentclass[a4paper,11pt]{article}
\usepackage{jheppub} % for details on the use of the package, please see the JINST-author-manual
\usepackage{lineno}
\usepackage{multirow}
\usepackage{graphicx}
\usepackage{amsmath}
\usepackage{amsfonts}
\usepackage{amssymb}
\usepackage{slashed} 
\usepackage{color}
\usepackage{hyperref}
\usepackage{units}
\usepackage{braket}
\usepackage{amsmath}
\nolinenumbers

\title{Probing primordial black holes from a first order phase transition through pulsar timing and gravitational wave signals}

 \author[1]{Jan Tristram Acu\~{n}a\note{Corresponding author.}}
 \author{and Po-Yan Tseng}
 \affiliation{Department of Physics, National Tsinghua University, 101 Kuang-Fu Rd., Hsinchu 300044,\\ Taiwan R.O.C.}
\emailAdd{jtacuna@gapp.nthu.edu.tw}
\emailAdd{pytseng@phys.nthu.edu.tw}

\abstract{In this work, we assess the sensitivity reach of pulsar timing array (PTA) measurements to probe pointlike primordial black holes (PBHs), with an extended mass distribution, which originate from collapsed Fermi balls that are formed through the aggregation of asymmetric U(1) dark fermions trapped within false vacuum bubbles during a dark first order phase transition (FOPT). The PBH formation scenario is mainly characterized by the dark asymmetry, strength of the FOPT, rate of FOPT, and the percolation temperature. Meanwhile, for PBH masses of interest lying within $10^{-10} M_\odot - 10^2 M_\odot$, the relevant signal for PTA measurements is the Doppler phase shift in the timing signal, due to the velocity change induced by transiting PBHs on pulsars. Taking the dark asymmetry parameter to be $10^{-4}$ and $10^{-5}$, we find that percolation temperatures within the $\unit[0.1 - 10]{keV}$ range, FOPT rates above $10^3$ times the Hubble parameter at percolation, and FOPT strengths within $10^{-6}-0.1$ can give rise to PBHs that can be probed by an SKA-like PTA observation. On the other hand, the accompanying gravitational wave (GW) signal from the FOPT can be used as a complementary probe, assuming that the peak frequency lies within the $\mathcal{O}(10^{-10})-\mathcal{O}(10^{-7})$ Hz range, and the peak GW abundance is above the peak-integrated sensitivity curves associated with pulsar timing observations that search for stochastic GWs. At the fundamental level, a quartic effective potential for a dark scalar field can trigger the FOPT. By performing a parameter scan, we obtained the class of effective potentials that lead to FOPT scenarios that can be probed by SKA through pulsar timing and GW observations.}

\begin{document}
\maketitle
\flushbottom

\section{Introduction}
\label{sec:Introduction}
Primordial black holes (PBHs) are hypothetical celestial objects formed in the early Universe and can serve as dark matter (DM) candidates~\cite{Hawking:1971ei,Chapline:1975ojl,Khlopov:2008qy,Carr:2016drx,Carr:2020gox,Carr:2020xqk,Green:2020jor}. Several production mechanisms for PBH formation have been proposed. PBHs may came from the collapse of overdense regions, which developed from primordial fluctuations after inflation~\cite{Carr:1974nx,Sasaki:2018dmp}. Other studies, \textit{e.g.} \cite{Hawking:1982ga,Kodama:1982sf,Moss:1994iq,Konoplich:1999qq}, invoke a cosmological first-order phase transition (FOPT), accumulating sufficient energy density within the Schwarzschild radius via bubble wall collisions, but such a mechanism is only efficient in producing supermassive PBHs heavier than $10^{10} M_\odot$. 

Recently, a novel scenario incorporating FOPT and dark sector particles has been proposed~\cite{Baker:2021nyl,Gross:2021qgx,Kawana:2021tde,Marfatia:2021hcp}, in which PBHs are formed from the collapse of macroscopic intermediate states called Fermi balls (FBs). Under this framework, which will be the main focus of this study, PBHs with sub-solar mass and a wide range of abundance can be generated.
We briefly describe the physical mechanism behind the formation of PBHs in this scenario.
The FOPT is induced by the finite temperature potential of a dark scalar $\phi$, when the dark sector temperature drops below the critical temperature $T_c$.
The dark fermions, denoted by $\chi$, need to be trapped in the false vacuum during FOPT, so that
the expanding true vacuum bubbles squeeze $\chi$ to form FBs.
A way to realize the trapping of dark fermions is by introducing a Yukawa interaction of the form $g_\chi \phi \bar{\chi}\chi$; when the $\phi$ gets a nonzero vacuum expectation value (VEV), $v_\phi$,
it lifts the mass difference of $\chi$ between the true and false vacua.
Further requiring $g_\chi v_\phi \gg T_c$ would trap $\chi$ particles in the false vacuum and form FBs.
Once the range of the Yukawa interaction, set roughly by the inverse of the thermal mass of $\phi$, becomes comparable with the separation distance between the bound $\chi$ particles in the FB, it triggers the collapse of FBs into PBHs.

Some constraints on the properties of PBHs can be obtained by considering limits from Hawking radiation and gravitational lensing~\cite{Carr:2020gox}.
For $M_{\rm PBH}\lesssim  10^{-16} M_\odot$, Hawking radiation becomes significant, and light particles from PBH evaporation contribute to cosmic ray fluxes in the present epoch. For $M_{\rm PBH}\lesssim 2.5\times 10^{-19} M_\odot$, the lifetime of such PBHs is shorter than the age of the Universe, but they can nevertheless inject entropy into the thermal bath through PBH evaporation, and may introduce distortions in the cosmic microwave background (CMB) and affect Big Bang nucleosynthesis (BBN). 
On the other hand, for $10^{-11} M_\odot \lesssim M_{\rm PBH}\lesssim 10\, M_\odot$, gravitational microlensing observations, \textit{e.g.} Subaru-HSC/OGLE-IV/EROS-2~\cite{Croon:2020ouk,Croon:2020wpr}, are able to set stringent limits on the PBH abundance.

In this work, we consider pulsar timing arrays (PTAs) as a complementary probe of PBHs within the window $10^{-10} M_\odot \lesssim M_{\rm PBH}\lesssim 10^{2}\, M_\odot$. 
A PTA consists of a catalog of millisecond pulsars (MSPs), which are astrophysically relevant due to their extremely accurate timing signals. This property of MSPs has been exploited to measure gravitational waves (GWs) in the nanohertz frequency band, by considering a PTA as an interferometer~\cite{Lommen:2015gbz,Tiburzi:2018txc}. 
However, the timing signal of an individual pulsar in a PTA can be affected by the gravitation of the surrounding PBHs, mainly through Doppler and Shapiro effects~\cite{Kashiyama:2012qz,Schutz:2016khr,Dror:2019twh,Ramani:2020hdo,Lee:2020wfn,Liu:2021zlt}.
The former happens when a PBH passes through the vicinity of either a pulsar or the Earth, and induces acceleration; as for the latter, the Shapiro effect is due to the time delay in the propagation of the signal when a PBH crosses the line-of-sight between a pulsar and the Earth. To obtain limits on PBH properties, we perform a Monte Carlo simulation to assign masses, positions, and velocities on PBHs, based on the PBH mass and phase space distributions. The amplitude of the phase shift, relative to the intrinsic pulsar timing signal can be used to compute a statistical quantity called the signal-to-noise ratio (SNR). For a single mock simulation, the representative SNR is given by the maximum SNR across all pulsars. The sensitivity of a PTA to a PBH formation scenario is then determined by checking if the maximum SNR is above a threshold value for 90\% of all mock simulations performed. 

In the case of PBHs formed during a dark FOPT in the early Universe, the PBH mass distribution and abundance can ultimately be derived from the finite-temperature quartic potential and particle content in the dark sector. The scalar potential determines the percolation temperature, FOPT strength, and FOPT rate, which directly impact the PBH mass distribution and abundance. In addition to the Doppler and Shapiro effects induced by PBHs, another useful probe of the FOPT are GWs with frequencies between $10^{-10}\,{\rm Hz}$ to $10^{-6}\,{\rm Hz}$, which can potentially be studied in future observations such as SKA~\cite{Janssen:2014dka}, THEIA~\cite{Theia:2017xtk}, and $\mu$Ares~\cite{Sesana:2019vho}.
Similar to the derivation of PTA limits, we calculate the associated SNR of a GW signal from a FOPT over the course of a 20-year observation period. As elaborated in detail in \cite{Schmitz:2020syl}, a convenient method to calculate the GW signal SNR is through the use of peak-integrated sensitivity curves (PISC). Under this framework, the relevant parameters are reduced into the peak GW abundance $\Omega_{\rm peak}h^2$ and peak frequency $f_s$; the PISC for a specific experiment and observation time simply corresponds to an SNR equal to unity. The FOPT parameters, to which PTA is sensitive, provide predictions for $\Omega_{\rm peak} h^2$ and $f_s$; the SNR can be easily calculated as the ratio of the predicted peak GW abundance to the PISC evaluated at the predicted peak GW frequency. 

This paper is organized as follows. We revisit the method of obtaining sensitivities on the phase shift of the pulsar timing signals via Doppler and Shapiro in Sec.~\ref{sec:PulsarTimingIntro}. In Sec.~\ref{sec:mono_PBH} we apply this method, as an illustrative example, to the case of monochromatic PBHs and obtain limits on the PBH fraction and PBH mass. Secs.~\ref{sec:PBHFOPT} and \ref{sec:MixedPBH-FB} cover mainly the PBH formation scenario via dark FOPT, which includes a discussion of the extended PBH mass function and mixed PBH-FB scenario, and the pulsar timing sensitivities projected on the FOPT parameter space.
We calculate the limits from GWs produced during a dark FOPT in Sec.~\ref{sec:GW}. We specify the quartic potential in Sec.~\ref{sec:potential}, which realizes the FOPT, and we use it to investigate the correlation between PTA and GW signals. Finally, we summarize the results in Sec.~\ref{sec:conclusion}. 
\section{Pulsar timing as probe of substructure}
\label{sec:PulsarTimingIntro}
In this section, we shall provide a detailed discussion, following closely references \cite{Ramani:2020hdo,Lee:2020wfn}, of pulsar timing as a way to determine the properties of a collection of transiting substructures. It is well-known that pulsars emit periodic signals, in the form of electromagnetic waves, that are extremely stable over extended time periods; the stability of pulsar signals is comparable with Earth-based atomic clocks. The pulsar signal arriving on Earth at time $t$ can be characterized by its phase $\phi(t)$, which is fitted with a polynomial function, referred to as the timing model, given by
\begin{eqnarray}
\label{phaseexpansion}\phi(t) = \phi_0 + \nu t + \frac{\dot{\nu}}{2} t^2,
\end{eqnarray}
where $\nu$ is the frequency of the emitted pulsar signal, and $\dot{\nu}/\nu$ is the spin-down rate. On the other hand, a transiting object in the vicinity of the pulsar, and/or close to the Earth-pulsar line of sight, can induce a phase shift to the pulsar emission. In Eq. (\ref{phaseexpansion}), the signal from such transiting objects will manifest starting at third-order; the additional contribution to the linear and quadratic terms can be reabsorbed into the fit to the pulsar period and spin-down rate \cite{Schutz:2016khr}. In this study, we shall focus on the \textit{Doppler phase shift} induced by the passage of compact objects; specifically for this case, we take the dominant signal to come from the velocity shift of the relevant pulsar, due to the acceleration on the pulsar caused by the transiting compact object. Another effect, known as the \textit{Shapiro time delay}, changes the propagation time of the signal along the line of sight due to the gravitational potential sourced by the transiting object. In the case of transiting PBHs, previous literature, \textit{e.g.} \cite{Schutz:2016khr,Dror:2019twh,Lee:2020wfn,Ramani:2020hdo}, has shown that the Doppler signal can be used to constrain monochromatic PBH masses as low as $\sim 10^{-11} - 10^{-10} M_\odot$, whereas the Shapiro signal loses sensitivity in that mass range. As for the frequency shift induced by the the Doppler and Shapiro effects, it can be shown that they are respectively given by \cite{Dror:2019twh}
\begin{eqnarray}
\left(\frac{\delta \nu}{\nu}\right)_D &=& \frac{1}{c}\hat{d}\cdot \int \vec{\nabla}\Phi~dt\\
\left(\frac{\delta \nu}{\nu}\right)_S &=& -\frac{2}{c^3}\vec{v}\cdot \int \vec{\nabla}\Phi~dz,
\end{eqnarray}
where $\Phi$ is the potential generated by the compact object, $\hat{d}$ is the unit vector pointing from the Earth to the pulsar, $\vec{v}$ is the velocity of the compact object. We take the assumption of a constant velocity trajectory, $\vec{r}=\vec{r}_0 + \vec{v}t$: this is a reasonable assumption since altering the velocity of the object requires it to be extremely close to the pulsar, and such strong-field events are rare \cite{Schutz:2016khr}.

For the Shapiro contribution to the frequency shift, the integration in $z$ is performed along the line-of-sight, and the gradient of the potential is understood to be evaluated at each point along the line-of-sight. Given a shift in frequency $\delta \nu (t)$, the corresponding phase shift is
\begin{eqnarray}
\delta\phi(t) = \int^t dt'~\delta \nu(t').
\end{eqnarray}
We then sift out the signal amplitude from the phase shift that cannot be reabsorbed into the intrinsic phase signal from the pulsar. This can be done by subtracting away
\begin{eqnarray}
\delta\phi_0 = \sum_{n=0}^2 f_n(t) \left[\frac{1}{T_{obs}}\int_0^{T_{obs}}dt'~\delta\phi(t') f_n(t')\right],\quad f_n(t) \equiv \sqrt{2n+1}P_n(2t/T_{obs}-1)
\end{eqnarray}
from $\delta\phi(t)$. For a collection of $N$ compact objects, each of which is labelled by an index $i = 1, ..., N$, the so-called residual signal amplitude $h_I(t)$ at the $I^{\text{th}}$ pulsar is
\begin{eqnarray}
\nonumber h_I(t) = \sum_{i=1}^{N}\delta\phi_{I,i}(t)-\delta\phi_{0,I}(t),\quad \delta\phi_{0,I} \equiv \sum_{n=0}^2 f_n(t) \left[\frac{1}{T_{obs}}\int_0^{T_{obs}}dt'~\sum_{i=1}^{N}\delta\phi_{I,i}(t') f_n(t')\right].\\
\end{eqnarray}
The signal-to-noise (SNR) ratio, a quantity which we will eventually use to assess the sensitivity of pulsar timing in probing the presence of transiting substructures, is evaluated using $h_I(t)$. This is given by
\begin{eqnarray}
\text{SNR}_I^2 = \frac{1}{\nu_I^2 t_{rms}^2\Delta t}\int_0^{T_{obs}}dt~h_I^2(t).
\end{eqnarray}
In the above expression, $\Delta t$ is the cadence of the pulsar timing measurement, which refers to the time interval between pulsar timing measurements; $T_{obs}$ is the observation time of the measurement; and $t_{rms}$ is the timing residual of the timing measurement, which refers to the uncertainty in the pulsar timing data. For pulsar timing observations comparable with the reach of SKA, we have $\Delta t = \unit[2]{weeks}, t_{rms} = \unit[50]{ns}, T_{obs} = \unit[20]{years}$ \cite{Rosado:2015epa}. The overall SNR is taken to be the maximum SNR across an array of pulsars, so that
\begin{eqnarray}
\text{SNR}^2 = \text{max}_{I} \left\{\text{SNR}_I^2\right\}.
\end{eqnarray}
The SNR, obtained from the residual signal, is completely determined from the positions and velocities of the compact objects, and the fixed positions of the pulsars. 
 
In determining the PTA sensitivity on the distribution of compact objects/substructures in the Galaxy, 
we shall adopt a frequentist, statistical approach where the initial positions and velocities of the compact objects are drawn from a phase space distribution, which can be factorized into a spatially-dependent part and a velocity-dependent part. We assume that the velocity distribution is a truncated Maxwell-Boltzmann distribution of the form
\begin{eqnarray}
f(\vec{v}) &=& N^{-1} \exp(-v^2/v_0^2)\Theta(v_{esc}-v),\\
N&\equiv&(\pi v_0^2)^{3/2}\left[\text{erf}(v_{esc}/v_0)-\frac{2}{\sqrt{\pi}}\left(\frac{v_{esc}}{v_0}\right)\exp(-v_{esc}^2/v_0^2)\right],
\end{eqnarray}
where $v_{rms} = \unit[325]{km/s}$, $v_{esc} = \unit[600]{km/s}$ \cite{Dror:2019twh}, and
\begin{eqnarray}
v_0 = \sqrt{\frac{2}{3}}v_{rms} \left[1-\frac{2}{3\sqrt{\pi}}\frac{z^3 \exp\left(-z^2\right)}{\text{erf}(z)-\frac{2}{\sqrt{\pi}}z\exp\left(-z^2\right)}\right]^{-1/2} \simeq \sqrt{\frac{2}{3}}v_{rms},\quad z \equiv \frac{v_{esc}}{v_0}.
\end{eqnarray}
Meanwhile, the spatial volume distribution of compact objects is assumed to be uniform within some simulation volume; in the Solar neighborhood, this can be taken to be a fraction of the local DM density $\rho_{DM} = \unit[0.46]{GeV/cm^3}$. Ultimately, this makes the SNR a random variable. In the absence of a signal, the SNR follows a one-sided Gaussian \cite{Dror:2019twh}; assuming that the SNRs from each pulsar are independent of each other, the joint probability density is just a product of these one-sided Gaussians. Given a threshold $p$-value, which gives the probability that the alternative hypothesis---in this case, the presence of a residual signal in the phase shift---occurred by random chance, the threshold SNR value, denoted by SNR$_*$, is
\begin{eqnarray}
\text{Pr}\left(\text{SNR}_1>\text{SNR}_* \cup ... \cup \text{SNR}_{N_{pul}}>\text{SNR}_*\right) = p,
\end{eqnarray}
where $N_{pul}$ is the number of pulsars in the array. Then we have
\begin{eqnarray}
\text{Pr}\left(\text{SNR}_1\leq\text{SNR}_* \cap ... \cap \text{SNR}_{N_{pul}}\leq\text{SNR}_*\right) = 1-p,
\end{eqnarray}
so that
\begin{eqnarray}
\text{Pr}\left(\text{SNR}\leq\text{SNR}_*\right)^{N_{pul}} = 1-p &\Rightarrow& \text{erf}(\text{SNR}_*/\sqrt{2})^{N_{pul}} = 1-p.
\end{eqnarray}
For $N_{pul}=200$ pulsars, the threshold SNR is $\text{SNR}_* = 3.66 \approx 4$ for $p=0.1$. We then construct a sufficiently large number of mock ``Universes," taken to be 1000, where we specify the initial positions and velocities of the compact objects, keeping the pulsar positions fixed; if a signal, \textit{i.e.} when the SNR is at least 4, exists in at least 90\% of randomly generated ``Universes," then PTA measurements are sensitive in probing the substructure under consideration.
\section{Sanity checks: pointlike, monochromatic PBH mass}
\label{sec:mono_PBH}
In the limit where the transiting object is pointlike, \textit{e.g.} in the case of primordial black holes (PBHs), and in the case where the velocity of the transiting object is regarded as a constant, the frequency shifts simplify into (see Appendix \ref{appendix:derivations} for more details)
\begin{eqnarray}
\left(\frac{\delta \nu}{\nu}\right)_D &=& \frac{G_N M_{PBH}}{v^2 c\tau_D}\frac{1}{\sqrt{1+x_D^2}}(x_D \hat{b}-\hat{v})\cdot\hat{d}\\
\left(\frac{\delta \nu}{\nu}\right)_S &=& \frac{4G_N M_{PBH}}{c^3 \tau_S}\frac{x_S}{1+x_S^2}.
\end{eqnarray}
Here, $\hat{b}$ is the unit vector pointing from the pulsar to the closest distance approach of the transiting object. In either case, $x$ refers to a rescaled time variable given by
\begin{eqnarray}
x_D = \frac{t - t_{0,D}}{\tau_D},\quad x_S = \frac{t - t_{0,S}}{\tau_S}.
\end{eqnarray}
The quantities $t_{0,D}$ and $t_{0,S}$ refer, respectively, to the time at which the object reaches the closest distance to the pulsar, and the closest distance to the Earth-pulsar line of sight. The time variable $\tau$, which exists for each signal, can be thought of as the characteristic width of the signal. For the Doppler signal, $\tau_D = b/v$, where $b$ is the closest distance between the pulsar and the transiting object. Meanwhile, for the Shapiro signal, $\tau_S = b_\perp/v_\perp$, where $b_\perp$ is the closest distance to the line of sight, and $v_\perp$ is the magnitude of the velocity component perpendicular to the line of sight. The phase shifts associated with the Doppler and Shapiro signals are then obtained by integrating the frequency shifts over time, and we get
\begin{eqnarray}
\label{PhiDoppler}\delta\phi_D(t) &=& \nu \frac{G_N M_{PBH}}{v^2 c}\left[(\hat{d}\cdot\hat{b})(1+x_D^2)^{1/2}-(\hat{d}\cdot\hat{v})\sinh^{-1} x_D\right]\\
\label{PhiShapiro}\delta\phi_S(t) &=& \nu \frac{2G_N M_{PBH}}{c^3}\ln(1+x_S^2).
\end{eqnarray}
In the above expressions, we ignore constants of integration, which will simply be reabsorbed by $\phi_0$ in Eq. (\ref{phaseexpansion}). 

For a given monochromatic PBH mass $M_{PBH}$, and PBH mass fraction $f_{PBH}$, one can obtain the local number density of PBHs, which will determine the size of the sampling volume where the locations of the PBHs will be drawn. Note that the shape of the sampling volume will be different for the Doppler signal and for the Shapiro signal. This can be understood by noting that the signal width for the Doppler signal depends on the closest distance to the pulsar, which makes a spherical volume centered at the pulsar a natural shape for the simulation volume. The radius of this spherical volume is
\begin{eqnarray}
\nonumber R_D = \left(\frac{3}{4\pi}\frac{N_{DM} M_{PBH}}{f_{PBH}\rho_{DM}}\right)^{1/3} &\simeq& \left(\unit[6.099 \times 10^{-6}]{kpc}\right)\left(\frac{N_{DM}}{100}\right)^{1/3}\left(\frac{M_{PBH}}{10^{-10} M_\odot}\right)^{1/3} \\
&\times& \left(\frac{\unit[0.4]{GeV/cm^3}}{\rho_{DM}}\right)^{1/3}\left(\frac{1}{f_{PBH}}\right)^{1/3}
\label{RDexp}
\end{eqnarray}
The radial position of a PBH within this volume can be specified by taking a random number $p_i$ from a uniform distribution on the unit interval, and this is given by
\begin{eqnarray}
r_i = p_i^{1/3} R_D.
\end{eqnarray}
Analogously, the Shapiro signal width depends on the closest distance to the Earth-pulsar line of sight; since the object must transit between the Earth and the pulsar, the natural shape of the simulation volume is a cylinder coaxial with the line of sight, with a length equal to the Earth-pulsar separation $L$. The radius of this cylindrical volume is
\begin{eqnarray}
\nonumber R_S &=& \left(\frac{1}{\pi L}\frac{N_{DM} M_{PBH}}{f_{PBH} \rho_{DM}}\right)^{1/2}\\
\nonumber&\simeq& \left(\unit[1.739 \times 10^{-8}]{kpc}\right)\left(\frac{N_{DM}}{100}\right)^{1/2}\left(\frac{M_{PBH}}{10^{-10} M_\odot}\right)^{1/2} \left(\frac{\unit[1]{kpc}}{L}\right)^{1/2} \\
&\times&\left(\frac{\unit[0.4]{GeV/cm^3}}{\rho_{DM}}\right)^{1/2} \left(\frac{1}{f_{PBH}}\right)^{1/2},
\label{RSexp}
\end{eqnarray}
so that a randomly assigned initial distance from the line of sight can be taken to be
\begin{eqnarray}
s_i = p_i^{1/2} R_S,
\end{eqnarray}
where, analogously, $p_i$ is taken from a uniform distribution on the unit interval.

\begin{figure}[t!]
    \centering
    \includegraphics[scale=0.42]{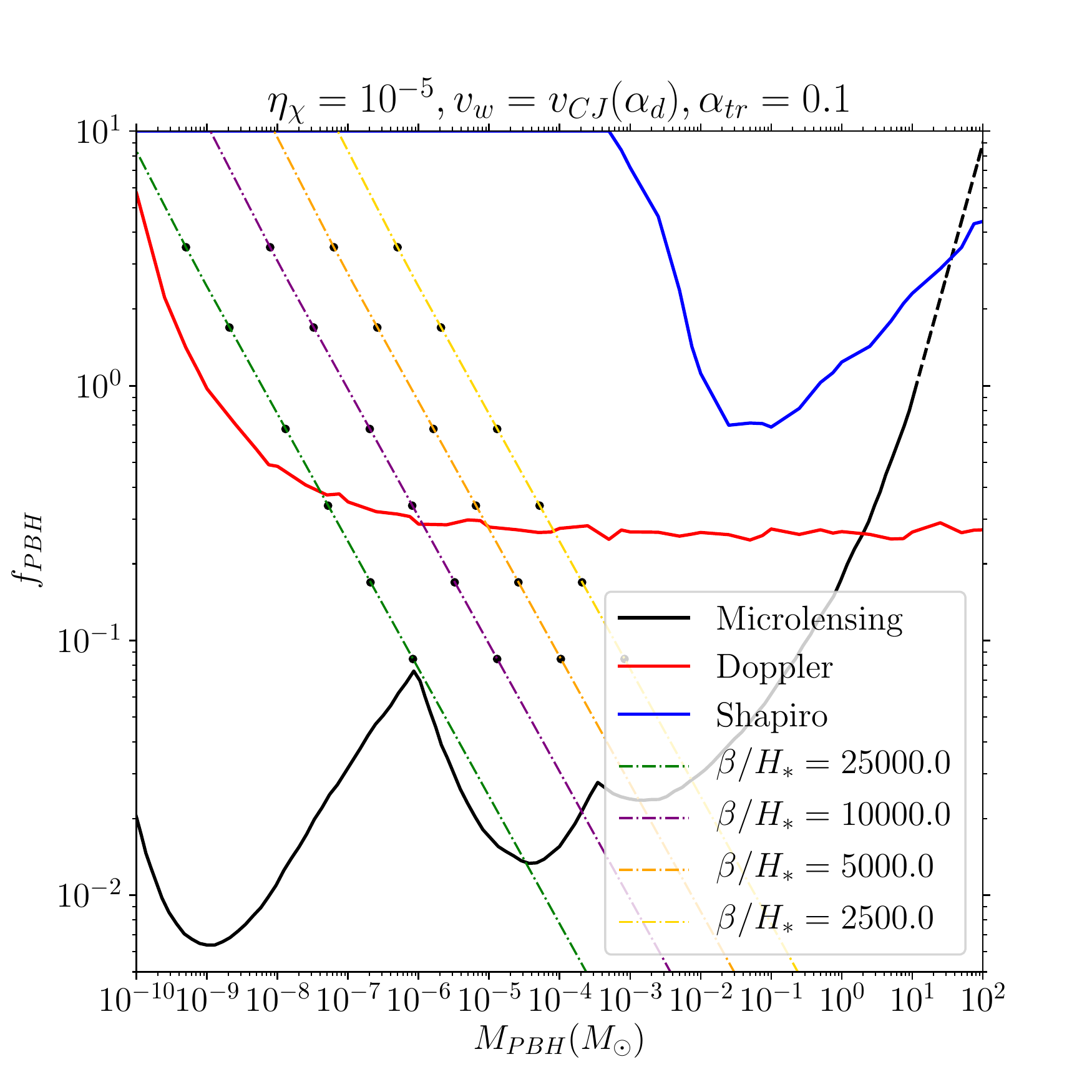}\includegraphics[scale=0.42]{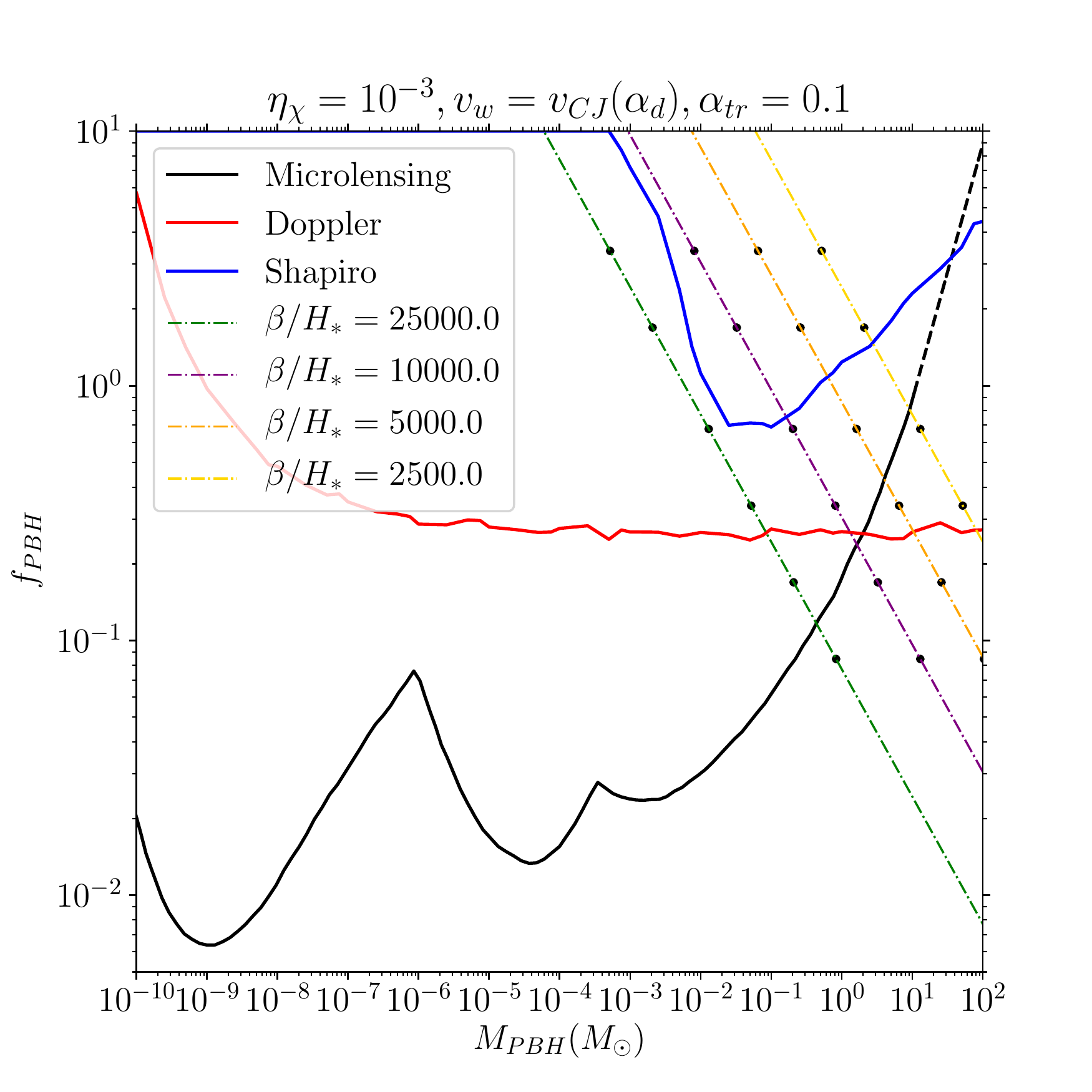}
    \caption{Sensitivity curves on the PBH parameter space, corresponding to Doppler signal, assuming SKA-like observational reach, and monochromatic PBH mass. We have included combined microlensing constraints for pointlike substructures from Subaru-HSC, OGLE-IV, and EROS-2 surveys, derived in \cite{Croon:2020wpr,Croon:2020ouk}. The dot dashed lines correspond to the projections of first order phase transition parameters (discussed in Section \ref{sec:PBHFOPT}) on the PBH fraction and the average PBH mass, for DM asymmetry parameters $\eta_\chi = 10^{-5}$ (left panel) and $\eta_\chi = 10^{-3}$ (right panel), and bubble wall velocity fixed to the Chapman-Jouguet value of $v_w(\alpha_d) \simeq 1$ for $\alpha_{tr}=0.1$ and $\alpha_d \sim 10^3$. In the left panel, the dot markers correspond to percolation temperature values 0.25, 0.50, 1.00, 2.00, 5.00, and \unit[10.0]{keV}; in the right panel, we have 2.5, 5.0, 10.0, 20.0, 50.0, and \unit[100.0]{eV}. Please refer to the text for more details. }
    \label{fig:PBHSensitivity}
\end{figure}
Fig. \ref{fig:PBHSensitivity} shows the result of our Monte Carlo simulation that we implemented using the procedure outlined in Section \ref{sec:PulsarTimingIntro}, where we fixed the number of samples within each simulation volume, associated with a single pulsar, to 100. Focusing on the Doppler signal sensitivity curve denoted by the red solid curve, we see that the PBH fraction stays flat for PBH masses of at least $\sim 10^{-6}M_\odot$, while lower PBH masses correspond to weaker sensitivity in the PBH fraction. This trend can be understood by considering the limit in which the PTA sensitivity is determined by a single PBH that yields the smallest signal width $\tau$, where we can ignore coherence/cancellation effects, which occur upon summing over the phase shifts from each PBH, due to the angular coefficients $\hat{d}\cdot\hat{b}$ and $\hat{d}\cdot\hat{v}$ in Eq. (\ref{PhiDoppler}). As discussed in \cite{Dror:2019twh}, for low PBH masses, the minimum sample signal width is much less than the observation time scale, which falls under the dynamical Doppler regime. In this case, the limit on the PBH fraction scales as the inverse of the PBH mass, mainly due to the fact that lighter PBHs must be sufficiently close to the pulsar to generate a large SNR. On the other hand, since the required simulation volume becomes larger for heavier PBHs, the typical distance of PBHs from pulsars would be comparable to or larger than the distance traversed by the PBH during the observation period, and this falls under the static Doppler regime. From dimensional analysis, the amplitude of the Doppler phase shift at third order in $t$ is proportional to $G_N M_{PBH}\nu T_{obs}^3/(v^2 c\tau_D^3)$; taking the smallest $\tau_D$ to be $b_{min}/v$, where $b_{min}$ is the smallest impact parameter among PBHs in the sample, one can show that $b_{min}$ scales as the PBH number density proportional to $(M_{PBH}/f_{PBH})^{1/3}$, and hence the dependence of the Doppler phase shift on the PBH mass disappears. As a side note, we observe from Fig. \ref{fig:PBHSensitivity} that gravitational microlensing provides the most stringent limits for the PBH mass range of interest; on the other hand, pulsar timing provides a complementary probe of these pointlike objects, and the sensitivity of pulsar timing can be improved by, \textit{e.g.}, increasing the observation time or increasing the number of pulsars in the catalog \cite{Dror:2019twh}. Furthermore, the Doppler signal sensitivity offers better discovery reach than microlensing for PBHs heavier than solar mass.

\begin{figure}[t]
    \centering
    \includegraphics[scale=0.38]{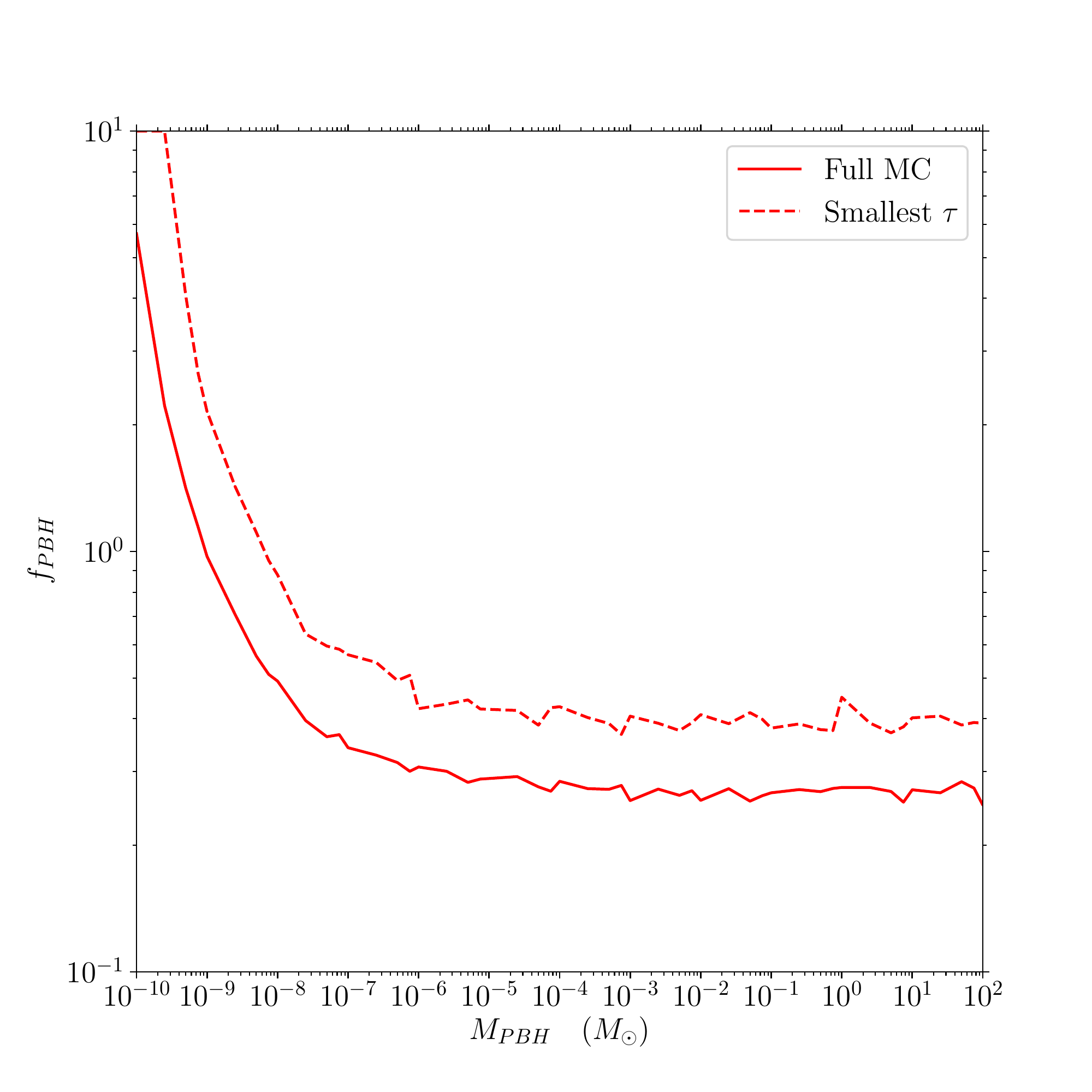}\includegraphics[scale=0.35]{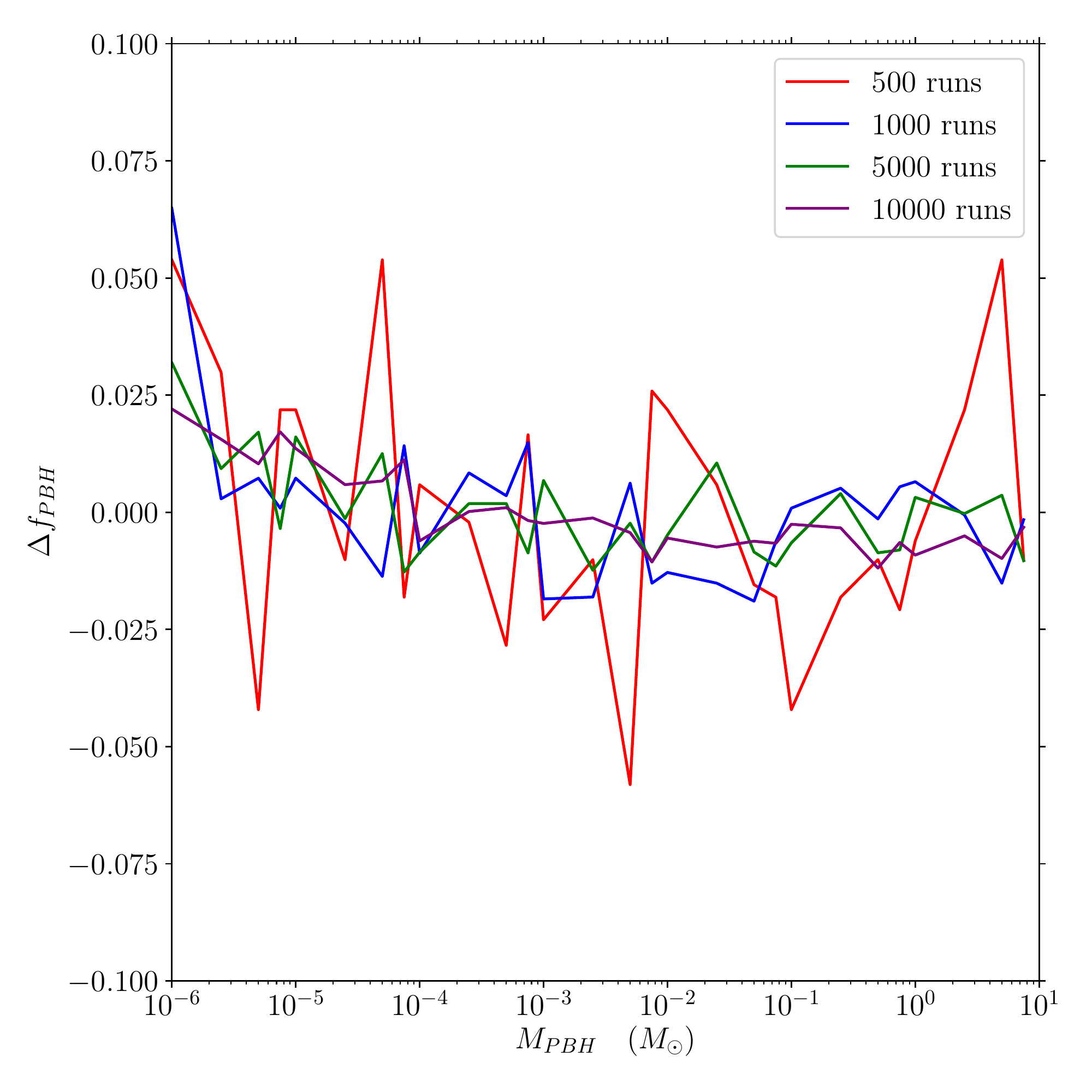}
   \caption{\label{fig:MCConsistencyChecks} The left panel is a comparison of the Doppler PTA sensitivity limits from a full MC simulation (solid), and the result of simply taking the PBH sample that yields the smallest signal width $\tau$ (dashed). We both see similar trends in the PBH fraction-PBH mass sensitivity. In the right panel, we show the fluctuation in the limit on the PBH fraction about the mean value, for different numbers of mock universes in the MC simulation. The variance in the PBH fraction decreases with increasing number of mock universes.}
\end{figure}

In the left panel of Fig. \ref{fig:MCConsistencyChecks}, we compare the result of a full MC simulation with simply taking the PBH sample, for each pulsar, with the smallest signal width. We expect fairly comparable results between them, based on the idea that the PBH with the smallest $\tau$ will provide the dominant contribution to the Doppler phase shift; in the actual plot, there appears to be an $\mathcal{O}(1)$ difference in the results because the full MC simulation considers a collection of PBHs, which can introduce some coherent enhancement in the amplitude of the Doppler phase shift. Meanwhile, we investigated the behavior of the fluctuation in the PBH fraction for $M_{PBH} \gtrsim 10^{-6}M_\odot$, when we change the number of mock universes in the MC simulation. Taking MC simulations with 500, 1000, 5000, and 10000 randomly generated universes, we found that the mean squared fluctuation $\sigma^2$ of $f_{PBH}$ within the considered mass range, decreases with increasing number of mock universes as listed in Table \ref{tab:runs}. This is an indication that the fluctuations in $f_{PBH}$ are a result of the random nature of the MC simulation itself. 
\begin{table}[t]
    \centering
    \begin{tabular}{|c|c|c|}
        \hline No. of runs & $\langle f_{PBH}\rangle$ & $\sigma^2$\\\hline
        500 & 0.2741 & $8.18 \times 10^{-4}$\\
        1000 & 0.2631 & $2.56 \times 10^{-4}$\\
        5000 & 0.2675 & $1.12 \times 10^{-4}$\\
        10000 & 0.2644 & $8.16 \times 10^{-5}$\\\hline
    \end{tabular}
    \caption{\label{tab:runs} Tabulated values of the average PBH fraction and the variance, in the mass range $10^{-6} M_\odot$ to $10^1 M_\odot$, of the PBH fractions plotted in the right panel of Fig. \ref{fig:MCConsistencyChecks}.}
\end{table}
\section{PBH from a dark first-order phase transition}
\label{sec:PBHFOPT}
We now consider a PBH formation mechanism that can eventually give rise to a PBH mass function, which goes beyond the case of a monochromatic PBH mass distribution. Following \cite{Lu:2022paj}, we consider the onset of a first-order phase transition (FOPT), during the radiation dominated era of the early Universe, \textit{within some dark sector}. This is triggered by the presence of a dark sector scalar field $\phi$ with an associated quartic potential given by $V_{eff}(\phi,T)$, where $T$ is the temperature of the dark radiation. In general, the visible and dark sector temperatures, which are respectively denoted by $T_{SM}$ and $T$, can be different; it is then convenient to introduce the temperature ratio parameter
\begin{eqnarray}
    \xi(t) \equiv \frac{T(t)}{T_{SM}(t)},
\end{eqnarray}
which can be time-dependent. Assuming that there is no exchange of energy and entropy between the two sectors, and mass thresholds are irrelevant, we shall take $\xi$ to be roughly constant in the relevant regime of interest. 

The VEV of the scalar field $\langle\phi\rangle$ is the order parameter characterizing the phase transition. At some critical temperature $T=T_c$, the VEV abruptly jumps from the false vacuum (FV) at $\langle\phi\rangle=0$, to the true vacuum (TV) at $\langle\phi\rangle = \phi_+$. In a FOPT, quantum tunneling allows the transition from the false to the true vacuum that are separated by a potential barrier. A bubble of the TV will form within a Hubble volume at the nucleation temperature $T = T_n < T_c$, corresponding to a time $t=t_n$ such that \cite{Wang:2020jrd} 
\begin{eqnarray}
1 = \int_{t_c}^{t_n}dt'~\frac{\Gamma(t')}{H^3(t')},
\end{eqnarray}
where $\Gamma(t)$ is the nucleation rate. The quantity $\Gamma(t)$ determines the rate at which TV bubbles form, within a comoving volume and within a comoving time. Given the three-dimensional instanton action $S_3(T)$, the nucleation rate is
\begin{eqnarray}
\label{S3NucRate}\Gamma(T) = T^4 \left[\frac{S_3(T)}{2\pi T}\right]^{3/2}\exp\left[-\frac{S_3(T)}{T}\right].
\end{eqnarray}
As time progresses, the dark sector plasma will be populated by TV bubbles, 
 and both phases coexist. For some comoving time $t$ larger than the comoving time $t_c$ at which the critical temperature is reached, the comoving radius of a TV bubble, formed at $t_i$, is 
\begin{eqnarray}
R_{TV}(t,t_i) = v_w \int_{t_i}^t \frac{dt'}{a(t')},
\end{eqnarray}
where $v_w$ is the bubble wall velocity. The exact dynamics of bubble wall expansion is beyond the scope of our study (see, \textit{e.g.} \cite{Moore:1995si,Moore:1995ua,Hindmarsh:2015qta,Laurent:2022jrs}), but we shall assume that the TV bubbles quickly reach their terminal velocity due to the friction from the surrounding plasma. The fraction of the Universe in the FV is then \cite{Turner:1992tz}
\begin{eqnarray}
\label{fFVdef}f_{FV}(t) &\equiv& \exp[-I(t)],\\
\label{argfFVdef}I(t) &\equiv& \frac{4\pi}{3}\int_{t_c}^t dt''~a^3(t'') R_{TV}^3(t,t'') \Gamma(T(t''))\simeq \frac{4\pi}{3}v_w^3\int_{t_c}^t dt''~\left(t-t''\right)^3 \Gamma(T(t'')).
\end{eqnarray}
In the last line, we note that the scale factor $a(t)$ goes as $t^{1/2}$ in the radiation dominated era, so one can show that the physical size of the TV, formed at $t_i$, is $a(t) R_{TV}(t,t_i) \simeq v_w (t-t_i)~\left[2/\left(1+\sqrt{t_i/t}\right)\right]$. Assuming that the phase transition is relatively short, we simply have $a(t) R_{TV}(t,t_i) \simeq v_w(t-t_i)$.

From the effective potential, we can define useful thermodynamic quantities. The free energy density difference between the FV and TV phases is
\begin{eqnarray}
\mathcal{F}(T) = \Delta V_{eff}(\phi_+(T),T);
\end{eqnarray}
the corresponding energy and entropy densities are respectively given by
\begin{eqnarray}
\epsilon_\phi(T) &=& \left(1-T\frac{\partial}{\partial T}\right)\Delta V_{eff}(\phi_+(T),T)\\
s_\phi(T) &=& -\frac{\partial \Delta V_{eff}(\phi_+(T),T)}{\partial T}.
\end{eqnarray}
The pressure is then
\begin{eqnarray}
p_\phi(\phi,T) = Ts_\phi - \epsilon_\phi = -\Delta V_{eff}(\phi_+(T),T).
\end{eqnarray}
Assuming that the scalar field can be described as a perfect fluid, the trace of the energy momentum tensor is
\begin{eqnarray}
\label{TraceAnomaly}\theta(T) \equiv \frac{1}{4}\left(\epsilon_\phi-3p_\phi\right) = \left(1-\frac{T}{4}\frac{\partial}{\partial T}\right)\Delta V_{eff}(\phi_+(T),T).
\end{eqnarray}
Given the trace of the energy momentum tensor, we can introduce 
\begin{eqnarray}
\label{AlphaTr}\alpha_{tr} \equiv \frac{\theta}{\rho_R},
\end{eqnarray}
which measures the \textit{strength of the phase transition}; here $\rho_R$ is the total radiation energy density, coming from the relativistic species in the dark and SM sectors, \textit{i.e.}
\begin{eqnarray}
\rho_R(t) &=& \rho_v(t)+\rho_{d,eq}(t),\\
\rho_v(t) = \frac{\pi^2}{30}g_{*\rho,v}(T_{SM})T_{SM}^4(t)&,&\quad \rho_{d,eq}(t) = \frac{\pi^2}{30}g_{*\rho,d}(T)T^4(t).
\end{eqnarray}
We can also define the strength of the phase transition normalized with respect to the dark sector radiation energy density, given by
\begin{eqnarray}
    \label{Def:adark}\alpha_d \equiv \frac{\theta}{\rho_{d,eq}},
\end{eqnarray}
which differs from $\alpha_{tr}$ by a factor
\begin{eqnarray}
    \frac{\rho_R}{\rho_{d,eq}} = \frac{g_{*\rho,v}(T/\xi)/\xi^4 + g_{*\rho,d}}{g_{*\rho,d}} \sim \frac{1}{\xi^4}.
\end{eqnarray}
In addition to the dark sector scalar field, we introduce a dark sector Dirac fermion $\chi$, possessing some global dark U(1) charge, and this may be regarded as a DM component. One can write down a renormalizable Lagrangian associated with $\phi$ and $\chi$, which respects this global U(1) symmetry, and it is given by
\begin{eqnarray}
\label{lagrangian}\mathcal{L} = \frac{1}{2}(\partial \phi)^2-V_{eff}(\phi, T)+\bar{\chi}\left(i\gamma^\mu \partial_\mu - m_\chi\right)\chi-g_\chi \phi \Bar{\chi}\chi.
\end{eqnarray}
In this study, we take the fermion mass parameter $m_\chi$ to be zero. Meanwhile, the trilinear Yukawa term serves a dual purpose of introducing an attractive interaction between dark fermions, and also provides additional mass to fermions within TV regions, equal to $g_\chi \phi_+(T)$, during the onset of the phase transition. We shall assume that during the relevant period of FOPT, there is no change in the effective relativistic degrees of freedom, and in the dark sector, we have
\begin{eqnarray}
    g_{*\rho,d} = 1 + \frac{7}{8}(4) = \frac{9}{2}.
\end{eqnarray} 
If we take the case where $g_\chi \phi_+(T) \gg T$, the DM particles lying in the FV do not have sufficient kinetic energy to penetrate through the bubble wall; this process is referred to as DM filtering \cite{Chway:2019kft}. As TV bubbles increase in size and population, the DM particles are trapped in pockets of FV bubbles that eventually shrink, forcing the DM particles to condense into Fermi balls (FBs); similar ideas were explored in, \textit{e.g.} \cite{Asadi:2021pwo}. Once the FBs form, no light degrees of freedom remain in the dark sector. Note that because the process of pair annihilation of dark fermions into scalars is quite efficient in depleting them, it is necessary to introduce an asymmetry in the number of DM particles relative to DM antiparticles; the asymmetry is denoted by the parameter
\begin{eqnarray}
    \eta_\chi \equiv \frac{n_\chi - n_{\bar{\chi}}}{s_v},
\end{eqnarray}
where 
\begin{eqnarray}
s_v(t_*) = \frac{2\pi^2}{45}g_{*s,v}(T_{SM}(t_*))T_{SM}^3(t_*) = \frac{2\pi^2}{45}g_{*s,v}(T(t_*)/\xi) \frac{T^3(t_*)}{\xi^3}
\end{eqnarray} 
is the entropy density of the visible sector, which is conserved assuming that no entropy nor energy exchange occurs between the visible and dark sector. The time $t = t_* > t_n$ refers to the \textit{percolation time}, where the FV filling fraction is taken to be $f_{FV}(t_*) = 0.29$ \cite{rintoul1997precise}. The corresponding number density of FV bubbles with radii lying between $R_r$ and $R_r + dR_r$ is \cite{Turner:1992tz,Kawana:2021tde}
\begin{eqnarray}
\label{dndR}\frac{dn}{dR_r} = \frac{\beta^4}{192v_w^4}I_*^4\exp(4\beta R_r/v_w)\exp[-I(t)]\left\{1-\exp[-I(t)]\right\},
\end{eqnarray}
where $\beta$, having the same units as the Hubble parameter, is roughly the inverse of the time duration of the FOPT, and $v_w$ is the bubble wall velocity. In obtaining Eq. (\ref{dndR}) we assumed that the TV nucleation rate can be written as 
\begin{eqnarray}
\label{GammaExp}\Gamma(t) = \Gamma_* \exp[\beta(t-t_*)],
\end{eqnarray}
which implies that the the FV filling fraction, $I(t)$, can be written as
\begin{eqnarray}
\label{gammastar_Istar} I(t) = \frac{8\pi v_w^3 \Gamma_*}{\beta^4}\exp[\beta(t-t_*)] \equiv I_* \exp[\beta(t-t_*)],~I_* = -\ln f_{FV}(t_*).
\end{eqnarray}
In obtaining Eq. (\ref{gammastar_Istar}), it is assumed that the saddle point approximation applies in evaluating Eq. (\ref{argfFVdef}). 

As the TV bubbles expand and increase in number during the FOPT, the regions containing the FV shrink and will eventually be completely surrounded by the walls of the TV bubbles. For this reason, the FV bubbles are not exactly spherical: quantitatively, the asphericity of a FV bubble with radius $R_*$ is captured by the parameter $A$, such that the volume of such a FV bubble is $A~4\pi R_*^3/3$. This parameter is directly tied to the FV filling fraction through the following normalization condition:
\begin{eqnarray}
f_{FV}(t_*) = \int dR_r~A\frac{4\pi R_r^3}{3}\frac{dn}{dR_r}(t_*).
\end{eqnarray}
The total U(1) charge in a FV bubble of radius $R_* = R_r\left(t_*\right)$ is given by \cite{Hong:2020est,Kawana:2021tde}
\begin{eqnarray}
\label{QFB}Q_{FB} = \frac{\eta_\chi s_{v}(t_*)}{f_{FV}(t_*)}A \frac{4\pi R_*^3}{3}
\end{eqnarray}
Subsequently, the FV bubble shrinks to form a FB containing the same amount of dark U(1) charge. While the individual dark fermions themselves are massless, the mass of the FB is a combination of the Fermi gas kinetic energy, the FV energy, and a negative contribution from the attractive Yukawa interaction. The total energy of a FB is a function of its radius, and the resulting FB configuration, at a given FB temperature, corresponds to the local energy minimum. In the limit where $\Delta V_{eff}(T_*) \gg T_*^4$, the FB mass is \cite{Kawana:2021tde,Marfatia:2021hcp}
\begin{eqnarray}
\label{MFB}M_{FB} = \left[12\pi^2 \Delta V_{eff}(T_*)\right]^{1/4} Q_{FB} \left[1+\mathcal{O}\left(T_*^2/\Delta V_{eff}^{1/2}\right)\right]
\end{eqnarray}
where $\Delta V_{eff}(T_*)$ is the energy difference between the false and true vacua, evaluated at the dark sector \textit{percolation temperature}, denoted by $T_* \equiv T(t_*)$. Using Eq. (\ref{QFB}), $Q_{FB}$ can be traded with $R_*$, so that a relation between the FB mass and $R_*$ can be established using Eq. (\ref{MFB}).

Due to the attractive Yukawa interaction, the FBs can collapse into PBHs. As the FB cools down, the range of the Yukawa interaction increases in the case where the dark scalar mass scales as the square of the temperature; at some collapse temperature $T_{cl}$, the Yukawa contribution to the FB energy becomes significantly large and the FB configuration ceases to be stable \cite{Kawana:2021tde,Huang:2022him}. Typically, since $T_{cl}$ is extremely close to the percolation temperature, and because of the mild temperature dependence of the FB mass according to Eq. (\ref{MFB}), the PBHs inherit the mass of the FB progenitor. From Eq. (\ref{dndR}), one can obtain the corresponding normalized probability distribution of PBH masses produced through this mechanism; this is given by
\begin{eqnarray}
P(M) &=& \frac{1}{f_{FV}(t_*)}\int dR_r~\delta\left(M-M_{PBH}(R_r)\right)\frac{dn}{dR_r}(t_*)~A~\frac{4\pi R_r^3}{3}\\
&=& \frac{1}{f_{FV}(t_*)}\frac{R_r}{3M}\frac{dn}{dR_r}(t_*)~A~\frac{4\pi R_r^3}{3}\\
\label{PMdist}&=& \frac{R_*}{3\left(12\pi^2 \Delta V_{eff}(T_*)\right)^{1/4}}\frac{dn}{dR_r}(t_*)~\frac{1}{\eta_\chi s_v(t_*)},
\end{eqnarray}
where $M$ is the mass of a PBH and
\begin{eqnarray}
\frac{dn}{dR_r}(t_*) = \frac{\beta^4}{192v_w^4}x^4\exp(-x)\left[1-\exp(-x)\right],~x(M) \equiv I_* \exp\left(\frac{\beta R_*(M)}{v_w}\right);
\end{eqnarray}
the relationship between $R_*$ and $M$ can be obtained from Eqs. (\ref{QFB}) and (\ref{MFB}), and identifying $M$ with $M_{FB}$. To obtain the SM entropy density, we use \cite{Drees:2015exa} to evaluate $g_{*s,v}(T_{SM}) \approx g_{*\rho,v}(T_{SM})$ for a given visible sector temperature $T_{SM}$.
\begin{figure}[t!]
    \centering
    \includegraphics[scale=0.42]{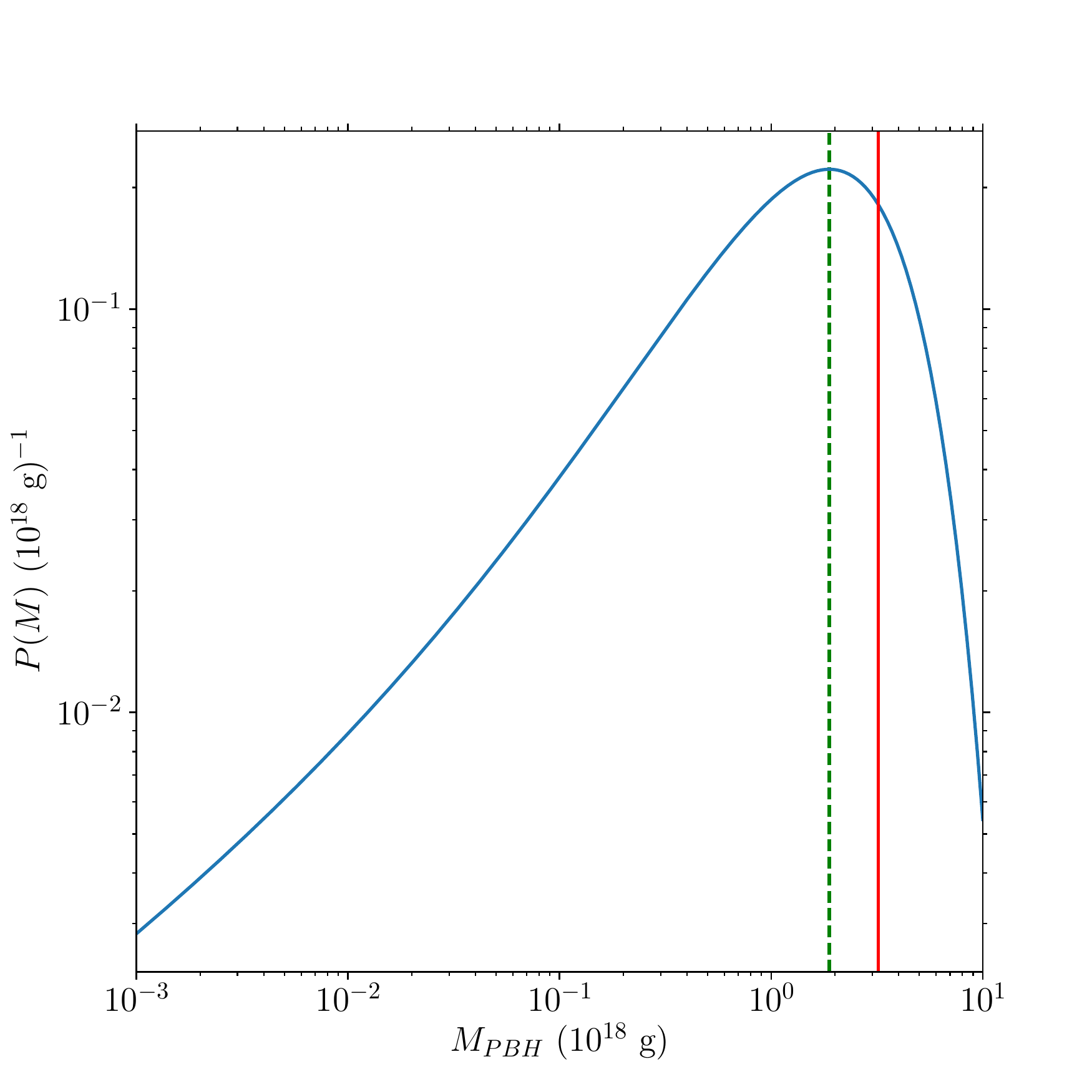}\includegraphics[scale=0.42]{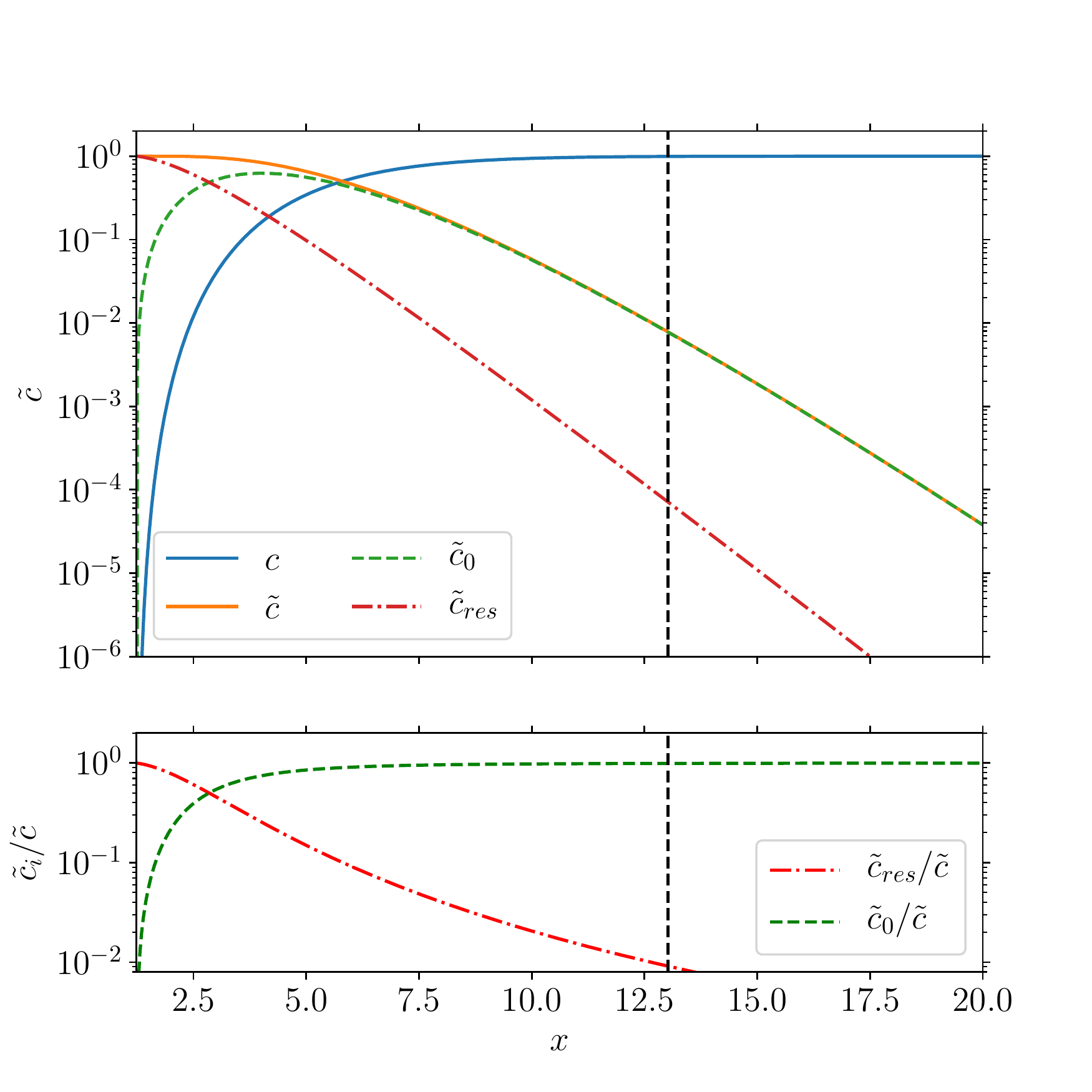}
    \caption{In the left panel we show the normalized probability distribution of PBH masses, in the case where the temperature of the Universe at percolation time is $T_* = \unit[200]{GeV}$, $g_* = 106.75$, $\eta_\chi = 10^{-3}$, $v_w = 0.2$, $\beta/H_* = 60$, and $\left[\Delta V_{eff}(T_*)\right]^{1/4} = \unit[100]{GeV}$. The vertical red, solid line and green, dashed line correspond to the average PBH mass $\langle M\rangle$ and PBH mass at the peak of the PBH mass distribution, respectively. Please refer to the main text for more details. The right panel shows the CDF $c$ associated with the PBH mass function, as a function of the universal dimensionless parameter $x(M)$. Note that this CDF only depends on $f_{FV}(t_*)$, and is unaffected by the choice of FOPT parameters; the FOPT parameters only dictate the form of $x(M)$. We also show the numerical result for $\tilde{c}$, $\tilde{c}_0$, and $\tilde{c}_{res}$ for $f_{FV}(t_*) = 0.29$; the black vertical line marks the position where $\exp(-xI_*)=10^{-7}$.}
    \label{fig:PBHMassFunc}
\end{figure}

In implementing the Monte Carlo simulation to calculate the PTA sensitivity curve, we need to construct sample universes by sprinkling each mock universe with PBHs, with the added feature that we have to assign masses to the PBHs in accordance with the distribution provided in Eq. (\ref{PMdist}). Sampling the distribution $P(M)$ can be done by first taking a random number $w$ from the unit interval $[0, 1)$, and then obtain $M$ such that $1-w = \tilde{c}(x(M))$, where $\tilde{c}(x(M))$ is the complement of the cumulative distribution function (CDF) $c(x(M))$ corresponding to $\tilde{P}(x)$, such that $P(M)~dM = \tilde{P}(x)~dx$, where
\begin{eqnarray}
\label{tildePx}\tilde{P}(x) \equiv \frac{A}{f_{FV}(t_*)}\frac{\pi}{144}x^3 \exp(-x)\left[1-\exp(-x)\right]\ln^3\left(\frac{x}{I_*}\right).
\end{eqnarray}
The CDF integral
\begin{eqnarray}
c(x(M)) = \frac{A}{f_{FV}(t_*)}\frac{\pi}{144}\int_{I_*}^{x(M)}dy~y^3 \exp(-y)[1-\exp(-y)]\ln^3\left(\frac{y}{I_*}\right)
\end{eqnarray}
is numerically stable when $x(M) \lesssim I_*$. For $x(M)\gg I_*$, the numerical evaluation of the complement function
\begin{eqnarray}
\tilde{c}(x(M)) \equiv \frac{A}{f_{FV}(t_*)}\frac{\pi}{144}\int_{x(M)}^{\infty}dy~y^3 \exp(-y)[1-\exp(-y)]\ln^3\left(\frac{y}{I_*}\right).
\end{eqnarray}
is more stable compared to the numerical integration of the CDF integral. By using the identity
\begin{eqnarray}
\int_x^\infty dy~F(y) \frac{d^3 G}{dy^3} = \left(FG''-F'G'+F''G\right)\Bigg\vert_{y=x}^{y=\infty} - \int_x^\infty dy~\frac{d^3 F}{dy^3}G(y),
\end{eqnarray} 
we can split $\tilde{c}(x(M))$ into
\begin{eqnarray}
\tilde{c}(x(M)) = \tilde{c}_0(x(M)) + \tilde{c}_{res}(x(M)),
\end{eqnarray}
where
\begin{eqnarray}
\nonumber\tilde{c}_0(x(M)) &\equiv& \frac{1}{8}\exp(-2x) x \ln(x/I_*)\\
\nonumber&\times&[-6 + 48 \exp(x) + 3 (-5 - 2 x + 8 \exp(x) (5 + x) )\ln(x/I_*)\\
 &+& (8 \exp(x) (6 + 3 x + x^2) - 2 (3 + 3 x + 2 x^2)) \ln^2(x/I_*)]\\
\nonumber \tilde{c}_{res}(x(M)) &=& \int_{x(M)}^\infty dy~\left[6 + 36 \ln(y/I_*) + 33 \ln^2(y/I_*) + 6 \ln^3(y/I_*)\right]\\
&\times&\exp(-y)\left[1-\frac{1}{8}\exp(-y)\right].
\end{eqnarray}
For $\exp(-x I_*) < 10^{-7}$, $\tilde{c}_{res}$ is at most 1\% of $\tilde{c}$, so we can simply drop $\tilde{c}_{res}$; this is demonstrated nicely in the right panel of Fig. \ref{fig:PBHMassFunc}.

Before resorting to a full Monte Carlo simulation, we can estimate the FOPT parameter region that is sensitive to pulsar timing array measurements, by first obtaining the dependence of the average PBH mass and PBH fraction on the FOPT parameters. In most cases, the temperature range at which the phase transition occurs is much smaller than $T_c$, so that we can perform a linear expansion of $\Delta V_{eff}$ around $T=T_c$, \textit{i.e.}
\begin{eqnarray}
\Delta V_{eff} \approx \epsilon_c\left(1-\frac{T}{T_c}\right),\quad \vert T-T_c\vert \ll T_c,
\end{eqnarray}
where $\epsilon_c \equiv \epsilon(T_c)$ can be interpreted as the latent heat that can be released to the plasma during the phase transition. The strength of the FOPT transition can be estimated as
\begin{eqnarray}
    \alpha_{tr} \approx \frac{\epsilon_c}{\rho_R(t_*)}\left(1-\frac{3}{4}\frac{T_*}{T_c}\right),
\end{eqnarray} 
and, in turn, we can express $\Delta V_{eff}(T_*)$ as
\begin{eqnarray}
    \Delta V_{eff}(T_*) \approx \alpha_{tr} \rho_R(t_*)\frac{\left(1-\frac{T_*}{T_c}\right)}{\left(1-\frac{3T_*}{4T_c}\right)}.
\end{eqnarray}
Then we can trade $\Delta V_{eff}(T_*)$ with $T_*$, $T_c$, and $\alpha_{tr}$, so that the set of relevant FOPT parameters are
\begin{eqnarray}
\nonumber\{\eta_\chi, T_*, \alpha_{tr}, T_c, \xi, \beta/H_*, v_w\}.
\end{eqnarray}
Furthermore, we can focus our attention to parameter regions which lead to reheat temperatures below the critical point, so that $\rho_R(T_c) - \rho_R(T_*) \gtrsim \epsilon_c$; in the limit $\vert T_c - T_*\vert \ll T_c$, this is equivalent to
\begin{eqnarray}
    \label{AlphaCrit}\alpha_{tr} \lesssim \left(\frac{T_c}{T_*}-1\right)\left[1+3\left(1-\frac{T_*}{T_c}\right)\right].
\end{eqnarray}
Typically, the lowest temperature $T_0$, at which the potential still develops a barrier between the false and true vacua, does not stray away from $T_c$. If we allow $T_* > T_0 \geq 0.8 T_c$, then this gives a maximum possible upper bound for $\alpha_{tr}$ of 0.4; in other words, the criterion that we do not reheat the dark plasma close to $T_c$ implies that $\alpha_{tr}$ is at most $\mathcal{O}(10^{-1})$.

To simplify our initial analysis, we shall take the following choices for the FOPT parameters:
\begin{eqnarray}
    \frac{T_*}{T_c} = 0.9,\quad \eta_\chi \in \{10^{-5}, 10^{-4}\},\quad \xi = 0.1.
\end{eqnarray}
For the temperature mismatch parameter $\xi$, the above choice can be shown to be consistent with constraints on $N_{eff}$; at the epoch of Big Bang nucleosynthesis (BBN), $\Delta N_{eff} \leq 0.4$ \cite{Iocco:2008va}. Firstly, if the phase transition had occurred before BBN, the absence of light degrees of freedom in the dark sector makes the BBN constraints on $N_{eff}$ irrelevant; otherwise, the contribution of the dark sector to $N_{eff}$ is
\begin{eqnarray}
    \Delta N_{eff}^{(BBN)} = \left(\frac{8}{7}\frac{\rho_R-\rho_\gamma}{\rho_\gamma}\right)_{BBN}\left(\frac{11}{4}\right)^{4/3} = 1.84 \times 10^{-4}\left(\frac{\xi}{0.1}\right)^4,
\end{eqnarray}
where $g_{*\rho,v}^{(BBN)} = 10.75$ and $g_{*\rho,d}^{(BBN)} = g_dtf = 4.5$. As for the bubble wall speed, we take
\begin{eqnarray}
    \label{vCJspeed} v_w = v_{CJ}(\alpha_d)=v_{CJ}(\alpha_{d}) \equiv \frac{1}{\sqrt{3}}\frac{1+\sqrt{2\alpha_{d}+3\alpha_{d}^2}}{1+\alpha_{d}}.
\end{eqnarray}
 Here we are assuming that the Chapman-Jouguet condition holds \cite{Steinhardt:1981ct}, where the bubble walls expand like spherical detonation waves; note that, since the formation of the bubbles occurred in the dark sector plasma, it is appropriate to evaluate the bubble wall velocity at $\alpha_d$, defined in Eq. (\ref{Def:adark}), rather than $\alpha_{tr}$. However, it has been pointed out by \cite{Laine:1993ey} that bubble wall expansion in the context of cosmological phase transitions can be characterized by a wider class of detonation scenarios. In \cite{Espinosa:2010hh} they provided contour plots for the bubble wall velocities in terms of the particle physics model parameters that trigger the FOPT. In this work, however, we are only interested in general trends in the limits on the FOPT parameter space, and we shall adopt the simplest approach where we stick with Eq. (\ref{vCJspeed}). Meanwhile, the above fixed choice for $T_*/T_c$ implies that $\alpha_{tr} \lesssim 0.14$; note that this upper limit on $\alpha_{tr}$ is a matter of choice based on Eq. (\ref{AlphaCrit}). Then the average PBH mass is given by:
\begin{eqnarray}
\langle M\rangle &=& \int dM~M P(M)\\
\nonumber &=& \left[12\pi^2 \Delta V_{eff}(T_*)\right]^{1/4}\frac{\eta_\chi s_v(t_*)}{H_*^3}\left(\frac{v_w}{\beta/H_*}\right)^3\\
&\times&\frac{(A~4\pi/3)^2}{192 f_{FV}^2(t_*)}\int_{I_*}^\infty dx~x^3 e^{-x}\left(1-e^{-x}\right)\ln^6\left(\frac{x}{I_*}\right)\\
\nonumber&\simeq& \left(4.07 \times 10^{-8} M_\odot\right)\left(\frac{10.63}{g_*}\right)^{1/4}\left(\frac{\unit[0.1]{MeV}}{T_*}\right)^2\left(\frac{\xi}{0.1}\right)^2\\
&\times&\left(\frac{\eta_\chi}{10^{-7}}\right)\left(\frac{v_w}{1}\right)^3\left(\frac{2.5 \times 10^2}{\beta/H_*}\right)^3\left(\frac{\alpha_{tr}}{0.1}\right)^{1/4}\left[\frac{\mathcal{F}(T_*/T_c)}{0.308}\right]^{1/4}
~\label{aveMPBHest}
\end{eqnarray}
where
\begin{eqnarray}
H_*^2 &=& \frac{8\pi}{3M_{Pl}^2}\rho_R,\quad \mathcal{F}(x) \equiv \frac{1-x}{1-\frac{3}{4}x}.
\end{eqnarray}
At percolation time, the total energy density of PBHs is
\begin{eqnarray}
\rho_{PBH,*} &=& \int dR~\left[12\pi^2 \Delta V_{eff}(T_*)\right]^{1/4}Q_{FB}(R)\frac{dn}{dR}\\
&=& \left[12\pi^2 \Delta V_{eff}(T_*)\right]^{1/4}\eta_\chi s_{v}(t_*).
\end{eqnarray}
Relative to the background cosmological cold DM (CDM) density, the total energy density of PBHs at percolation time is
\begin{eqnarray}
\omega_{PBH,*} &\equiv& \frac{\rho_{PBH,*}}{\rho_{cdm,*}} = \frac{s_{v,0}}{\rho_{cdm,0} s_v(t_*)}\int dR~\left[12\pi^2 \Delta V_{eff}(T_*)\right]^{1/4}Q_{FB}(R)\frac{dn}{dR}\\
\nonumber&=& \frac{\left[12\pi^2 \Delta V_{eff}(T_*)\right]^{1/4}\eta_\chi s_{v,0}}{\rho_{cdm,0}} \simeq 0.434\left(\frac{\alpha_{tr}}{0.1}\right)^{1/4}\left(\frac{g_*}{10.63}\right)^{1/4}\left(\frac{T_*}{\unit[0.1]{MeV}}\right)\\
&~&~~~~~~~~~~~~~~~~~~~~~~~~~~~~~~~~~\times\left(\frac{0.1}{\xi}\right)\left(\frac{\eta_\chi}{10^{-7}}\right)\left[\frac{\mathcal{F}(T_*/T_c)}{0.308}\right]^{1/4},
\label{fPBHest}
\end{eqnarray}
where $s_{v,0}$ is the present day entropy density in the visible sector, and $\rho_{cdm,0} \simeq \unit[1.26 \times 10^{-6}]{GeV/cm^3}$ is the current background cosmological CDM density. If we regard PBHs as a component of CDM, we can assume that the local PBH density relative to the local DM density, at present time, is simply given by
\begin{eqnarray}
f_{PBH} \equiv \frac{\rho_{PBH,\odot}}{\rho_{DM,\odot}} = \omega_{PBH,*}.
\end{eqnarray}
Using Eqs. (\ref{aveMPBHest}) and (\ref{fPBHest}), we can then obtain the projection on the $f_{PBH}-M_{PBH}$ plane, for certain choices of the FOPT parameters. This is shown in Fig. \ref{fig:PBHSensitivity}, where we project out contours of constant $\beta/H_* = 2.5 \times 10^3, 5.0 \times 10^3, 10^4, $ and $2.5 \times 10^4$, for DM asymmetry parameters $\eta_\chi = 10^{-5}$ and $10^{-3}$. Note that larger asymmetry parameters lead to larger dark U(1) charges in FBs, implying heavier PBHs. However, to maintain the same PBH fraction, an increase in $\eta_\chi$ must be compensated by a decrease in the percolation temperature. The dots in Fig. \ref{fig:PBHSensitivity} correspond to different values of the temperature at percolation $T_*$; on a contour with fixed $\beta/H_*$, the percolation temperature decreases in the direction of increasing average PBH mass, since $\langle M_{PBH}\rangle$ scales with the percolation temperature as $T_*^{-2}$. Meanwhile, smaller values of $\beta/H_*$ lead to larger FV bubble radii, which then correspond to heavier PBH masses; this trend can also be seen in the contours in Fig. \ref{fig:PBHSensitivity}. The locations of the dots in the contours of Fig. \ref{fig:PBHSensitivity}, relative to the PTA sensitivity curve obtained by assuming a monochromatic PBH mass distribution, indicate that PTA observations with similar properties as SKA would be sensitive to PBHs formed through a FOPT, with an associated percolation temperature $T_*$ lying around $\mathcal{O}(1)$ keV, with a FOPT rate around $\mathcal{O}(10^3-10^4) H_*$, assuming that the DM asymmetry is $10^{-5}$; for $\eta_\chi = 10^{-3}$, $T_*$ must be around $\mathcal{O}(10)$ eV.\\ 
\begin{figure}[t]
    \centering
    \includegraphics[scale=0.43]{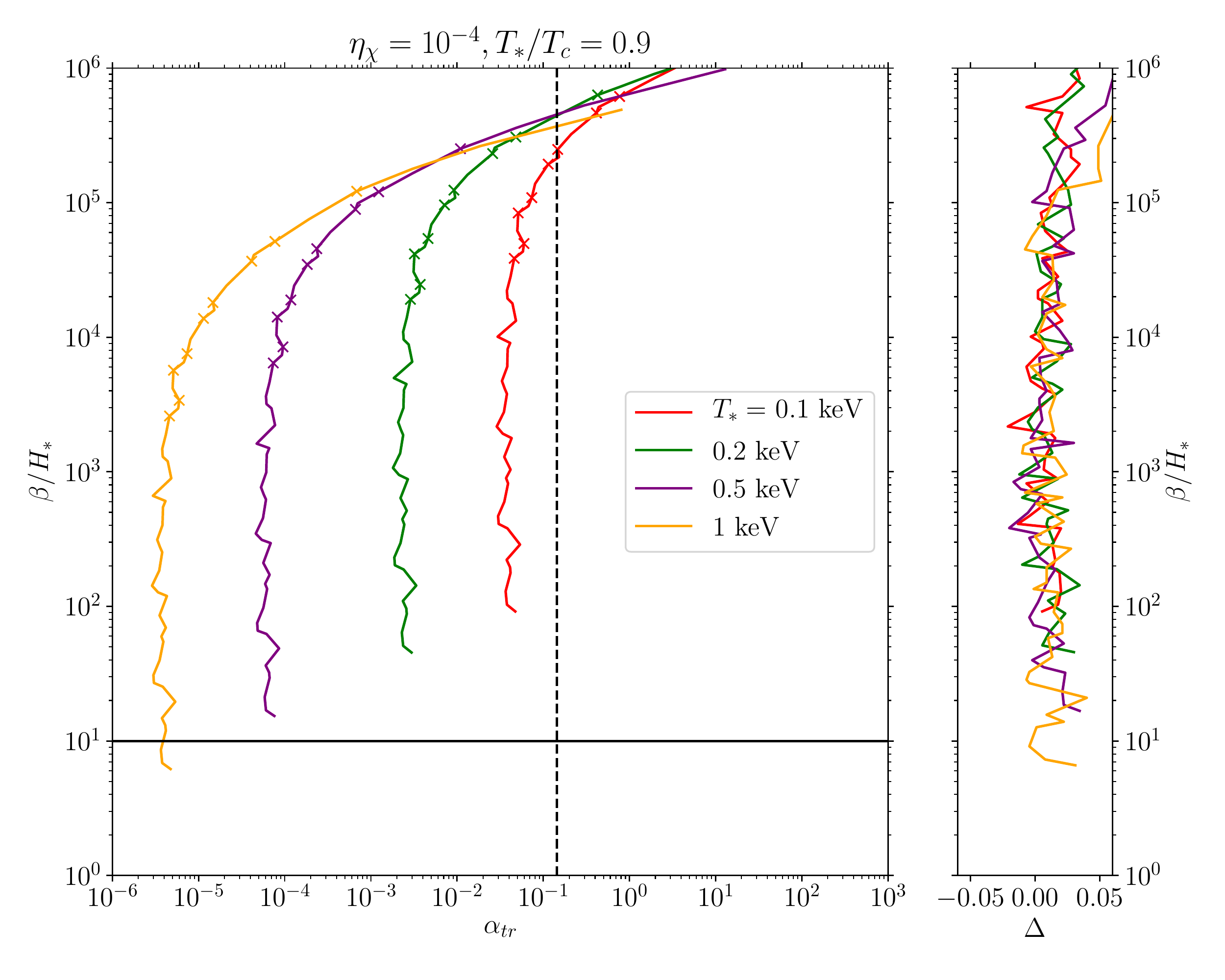}
    \includegraphics[scale=0.43]{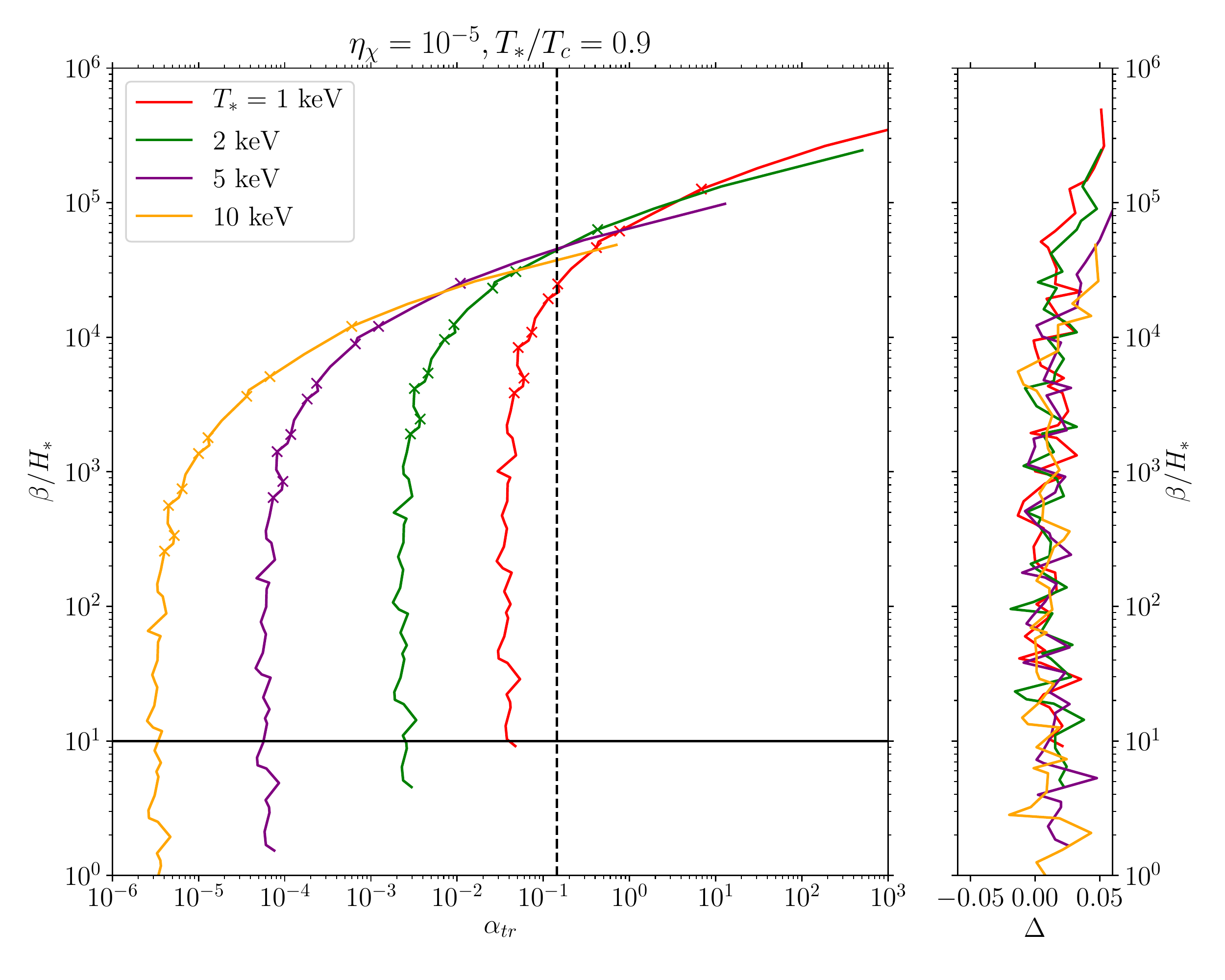}
    \caption{\label{fig:PBHFOPTSens} We show plots of the projection on the $\alpha_{tr}-\beta/H_*$ plane, of the PTA sensitivity curves for monochromatic PBH masses, where we match the monochromatic PBH mass to $\langle M\rangle$. We have appropriately chosen the dark sector percolation temperatures for $\eta_\chi = 10^{-4}$ and $10^{-5}$. The solid curves were obtained using Eqs. (\ref{alph_est}) and (\ref{boverH_est}). The cross marks correspond to average PBH masses $\langle M\rangle = 10^{-5}, 5 \times 10^{-6}, 10^{-6}, 5 \times 10^{-7}, 10^{-7}, 5 \times 10^{-8}, 10^{-8}, 5 \times 10^{-9}$, and $10^{-9} M_\odot$. For each panel, the quantity $\Delta$ measures the relative deviation from 0.9 of the fraction of simulated universes that provide SNR above threshold, obtained from an actual Monte Carlo simulation where the estimated FOPT parameter points were taken as input.} 
\end{figure}
~\\
Alternatively, we can invert Eqs. (\ref{aveMPBHest}) and (\ref{fPBHest}) to determine the FOPT parameters in terms of $\langle M_{PBH}\rangle$ and $\omega_{PBH,*}$. We have
\begin{eqnarray}
\nonumber\alpha_{tr} &=& \frac{\omega_{PBH*}^4}{12\pi^2 \mathcal{F}(T_*/T_c)}\left(\frac{\rho_{cdm,0}}{\eta_\chi s_{v,0}\rho_{R*}^{1/4}}\right)^4 \\
&\simeq& 0.16 \left(\frac{\omega_{PBH*}}{0.49}\right)^4\left(\frac{10^{-7}}{\eta_\chi}\right)^4\left(\frac{\unit[0.1]{MeV}}{T_*}\right)\left[\frac{0.308}{\mathcal{F}(T_*/T_c)}\right],
~\label{alph_est}\\
\frac{\beta}{H_*} &=& \left[12\pi^2 \mathcal{F}(T_*/T_c)\right]^{1/12} \alpha_{tr}^{1/12}\left(\frac{\eta_\chi s_{v*}\rho_{R*}^{1/4}}{H_*^3}\right)^{1/3} v_w(\alpha_{tr})\left(\frac{c_0}{\langle M\rangle}\right)^{1/3}\\
\nonumber &\simeq& \left(4.74 \times 10^2\right)\left(\frac{\alpha_{tr}}{0.16}\right)^{1/12}\left(\frac{7.5 \times 10^{-9}M_\odot}{\langle M\rangle}\right)^{1/3}\left(\frac{\eta_\chi}{10^{-7}}\right)^{1/3}\left(\frac{v_w}{1}\right)\left(\frac{\unit[0.1]{MeV}}{T_*}\right)^{2/3}\\
&\times&\left(\frac{10.63}{g_*}\right)^{1/12}\left[\frac{\mathcal{F}(T_*/T_c)}{0.308}\right]^{1/12},
~\label{boverH_est}
\end{eqnarray}
where $c_0 \equiv \frac{1}{192}(A~4\pi/3/f_{FV}(t_*))^2\int_{I_*}^\infty dx~x^3 e^{-x}\left(1-e^{-x}\right)\ln^6\left(x/I_*\right) \simeq 70$. For fixed percolation temperature $T_*$ and DM asymmetry $\eta_\chi$, the estimates of the FOPT parameters $\alpha_{tr}$ and $\beta/H_*$ given in Eqs. (\ref{alph_est}) and (\ref{boverH_est}) are useful to roughly pinpoint the parameter region where pulsar timing is expected to be sensitive, even without performing a full Monte Carlo simulation of the PBHs. In Fig. \ref{fig:PBHFOPTSens} we display the contours of constant percolation temperature $T_*$ on the $\alpha_{tr}-\beta/H_*$ plane, obtained from Eqs. (\ref{alph_est}) and (\ref{boverH_est}), which correspond to the expected sensitivity reach of an SKA-like PTA observation to measure Doppler signals produced by monochromatic PBHs of mass $\langle M\rangle$ and fraction $\omega_{PBH*}$. For a given $T_*$, the FOPT parameter points below the projected sensitivity curve will give rise to PBHs that provide a maximum SNR above threshold for the Doppler signal. The vertical dashed line corresponds to the maximum $\alpha_{tr}$, for a fixed $T_*/T_c$, beyond which the dark sector plasma will be reheated to the critical temperature and the false and true vacua will undergo a prolonged period of phase coexistence \cite{Megevand:2007sv}. Meanwhile, the horizontal line at $\beta/H_* = 10$ marks a convenient lower bound of phase transition rates that are faster than the Hubble expansion at percolation. 

To compare the results with a full Monte Carlo simulation, we can instead take the estimates for the viable FOPT parameters from Fig. \ref{fig:PBHFOPTSens}, and determine the fraction of mock universes $f$ that produce a maximum SNR above threshold. We expect this fraction to be close to 0.9, so we define a parameter $\Delta \equiv f/0.9 - 1$ which quantifies how good the estimates are in representing the PTA sensitivity limit in the FOPT parameter space. Looking at right hand side of each panel in Fig. \ref{fig:PBHFOPTSens}, we find that $\vert\Delta\vert$ is at most 5\%; this rather satisfactory agreement between the estimates and the full MC result can be attributed to the shape of the PBH mass distribution being peaked around $\langle M\rangle$, so the PBH mass distribution can roughly be characterized by a monochromatic distribution centered at $\langle M\rangle$. Notice also that the projected curves are similar in shape as the corresponding PTA sensitivity curves: the branch where the PBH fraction is fixed is mapped into a line of constant $\alpha_{tr}$, while the low-mass branch, where the PBH fraction becomes larger, is mapped into a rising branch in the $\alpha_{tr}-\beta/H_*$ plane. We see that the projected curves, for $\eta_\chi = 10^{-5}~(10^{-4})$, fall on the desired FOPT parameter region, \textit{i.e.} $\alpha_{tr}$ less than the maximum value to avoid reheating the dark plasma to $T=T_c$, and for $\beta/H_* > 10$, for percolation temperatures ranging from \unit[1-10]{keV} (\unit[0.1-1]{keV}).
\section{Mixed PBH-FB scenario}
\label{sec:MixedPBH-FB}
%\textcolor{red}{(JTA: I am starting to get a little suspicious about the feasibility of this scenario. I think that for any nonzero Yukawa coupling, the temperature at collapse $T_\phi$ can be made arbitrarily close to the destabilization temperature, such that the range of the Yukawa interaction is long enough... However, on the other hand, once Fermi balls are formed, their temperature evolution is dictated by Stefan-Boltzmann and \textbf{not through cosmological evolution}; is this the way out then? If the cooling time from $T_*$ to $T_\phi$ is longer than, e.g. the age of the Universe, then FB will not collapse to PBHs. The cooling mechanism is model-dependent: if the rate is dependent on the Yukawa coupling, this might be problematic in the context of realizing the filtered DM scenario; otherwise, if one only needs a small portal coupling to SM particles, then a mixed PBH-FB scenario can be realized then)}
In the previous section, we considered the scenario in which all of the FBs collapse into PBHs. We assume that the intermediate step of FB formation occurs, so we require that, at percolation time, the range of the Yukawa interaction $L_\phi \simeq M_\phi^{-1}(T_*) $ is less than the average interparticle distance, which scales as $R_*/Q_{FB}^{1/3}$, between $\chi$ within the FBs; otherwise, the FV bubbles containing trapped DM particles will immediately collapse into PBHs \cite{Kawana:2021tde}. If FBs are formed, a necessary condition for the collapse to occur is a mechanism that allows FBs to cool down, which can be achieved through the emission of a light particle \cite{Kawana:2021tde,Huang:2022him}, or the emission of neutrinos that can carry away energy from the FB. The latter process, mentioned in \cite{Witten:1984rs}, may be realized by assuming a portal interaction between the scalar $\phi$, introduced in Eq. (\ref{lagrangian}), and the Higgs. If the emitted particles are relativistic, then the energy loss rate scales as $T^4$ according to the Stefan-Boltzmann law; furthermore, if the cooling rate is faster than the Hubble expansion, this ensures that the FB will inevitably collapse into PBHs \cite{Kawana:2021tde,Bai:2018dxf} (see also \cite{Lu:2022jnp} for a recent discussion on blackbody and evaporative cooling). On the other hand, this criterion is dependent on the model which describes the DM interaction with light particles; if the couplings are sufficiently weak such that, \textit{e.g.} the cooling time is longer than the age of the Universe, then some of the FBs may not collapse into PBHs. In addition, the mixed PBH-FB scenario may be realized by noting that FBs lying on the low mass tail of the FB distribution will cool down faster than heavier FBs, and form PBHs. Hence, a possible final outcome might involve a situation where a fraction $r$ of substructures from FOPT come in the form of FBs, and the remainder, $(1-r)$, in PBHs. The exact fraction $r$ appears to be dependent on the underlying mechanism or DM microphysics, so we shall treat $r$ as a model independent parameter. The distributions for the FB radii and masses remain unchanged, and can be sampled by picking a random value of $x$ using the distribution in Eq. (\ref{tildePx}); the additional feature in our setup is the probability that a simulated substructure is a FB, given by $r$. Accounting for the finite size of the FB, the phase signal is multiplied by a form factor as prescribed in Appendix \ref{appendix:derivations}. For a spherically symmetric, constant volume density FB \cite{Marfatia:2021twj} with radius $R_{FB}(M)$, the form factor can be obtained using Eq. (\ref{ffac}), and we find that
\begin{eqnarray}
\label{constdensffac}\mathcal{F}(M,x) = \frac{3x}{R_{FB}(M)}\int_0^\infty \frac{\sin u - u \cos u}{u^3}J_1\left(\frac{u x}{R_{FB}}\right)~du.
\end{eqnarray}
For a given substructure mass $M$, the form factor in Eq. (\ref{constdensffac}) must be evaluated at the instant where the object is closest to the pulsar (in the case of Doppler), or where the object is closest to the line of sight (in the case of Shapiro). Note also that Eq. (\ref{constdensffac}) tends to unity in the limit where $x \gg R_{FB}$. Hence, to determine whether a Doppler signal can be used to distinguish between FBs and PBHs in this scenario, we can compare the typical distance between the object and the pulsar, given in Eq. (\ref{RDexp}), versus the average radius of FV bubbles, $\langle R\rangle$, which is an upper bound for the average radius of FBs, $\langle R_{FB}\rangle$. The average FV bubble radius is
\begin{eqnarray}
\nonumber\langle R\rangle &=& \frac{1}{f_{FV}(t_*)}\int dR_r~A\frac{4\pi R_r^3}{3}\frac{dn}{dR_r} \\
&=& \left(\frac{v_w}{\beta}\right)\frac{A~4\pi/3}{192 f_{FV}(t_*)}\int_{I_*}^\infty dx~x^3 \exp(-x)\left[1-\exp(-x)\right]\ln^4\left(\frac{x}{I_*}\right),
\end{eqnarray}
so that we have
\begin{eqnarray}
\frac{R_D}{\langle R_{FB}\rangle} \gtrsim \left(2.47 \times 10^{7}\right) \left(\frac{N_{DM}}{100}\right)^{1/3}\left(\frac{g_*}{10.63}\right)^{1/3}\left(\frac{T_*}{\unit[0.1]{MeV}}\right)\left(\frac{0.1}{\xi}\right).
\end{eqnarray}
The above estimate tells us that the typical size of FBs is much smaller than the typical pulsar-object distance, so a FB is effectively indistinguishable from a PBH from the point of view of a pulsar. Similarly, for the Shapiro signal, the relevant length scale is $R_S$ given by Eq. (\ref{RSexp}); compared to the typical FB size, we have
\begin{eqnarray}
\nonumber\frac{R_S}{\langle R_{FB}\rangle} &\gtrsim& \left(5.11 \times 10^{6}\right)\left(\frac{N_{DM}}{100}\right)^{1/2}\left(\frac{v_w}{1}\right)^{1/2}\\
&\times&\left(\frac{300}{\beta/H_*}\right)^{1/2}\left(\frac{g_*}{10.63}\right)^{1/4}\left(\frac{T_*}{\unit[0.1]{MeV}}\right)^{1/2}\left(\frac{0.1}{\xi}\right)^{1/2}.
\end{eqnarray}
Hence, even the Shapiro signal cannot discriminate between a pointlike object and a finite size FB. 
\section{Gravitational wave production}
\label{sec:GW}
During a FOPT, gravitational waves (GWs) can be produced, and these can serve as complementary signals that we can utilize to probe FOPT scenarios in the early Universe. It has been shown in \cite{Hindmarsh:2015qta} that the dominant contribution to the GW abundance from a FOPT comes from sound waves. The signal spectrum is given by (see \cite{Schmitz:2020syl} and references therein)
\begin{eqnarray}
\Omega_s (f) h^2 &=& \Omega_s^{peak} h^2~\mathcal{S}_s(f,f_s),\\
\label{Omh2PeakSound}\Omega_s^{peak}h^2 &\simeq& 2.65 \times 10^{-6}\left(\frac{v_w}{\beta/H_*}\right)\left[\frac{100}{g_{*\rho,v}(T_*)}\right]^{1/3}\left(\frac{\kappa_s \alpha_{tr}}{1+\alpha_{tr}}\right)^2\left(1+\frac{g_{*\rho,d}}{g_{*\rho,v}}\xi^4\right),\\
\label{SigSpexSound}\mathcal{S}_s(f,f_s) &=& \left(\frac{f}{f_s}\right)^3\left[\frac{7}{4+3(f/f_s)^2}\right]^{7/2},\\
\nonumber f_s &\simeq& \unit[1.9\times 10^{-2}]{mHz}\left[\frac{g_{*v}(T_*/\xi)}{100}\right]^{1/6}\left(\frac{T_*}{\unit[100]{GeV}}\right)\left(\frac{\beta/H_*}{v_w}\right)\left(1+\frac{g_{*\rho,d}}{g_{*\rho,v}}\xi^4\right)^{1/2}\frac{1}{\xi}.\\
\label{PeakFreqSound}
\end{eqnarray}
%\textcolor{red}{(JTA: This prescription for $\kappa$ might no longer work, since we are in the regime of a runaway phase transition because the bubble wall velocities are very close to the speed of light. We need to use Eqs. (2.7) and (2.9) of \cite{Schmitz:2020syl}; in Eq. (2.7) we have $\alpha_\infty$, which requires us to have the fermion masses in the false and true vacua, and therefore the Yukawa coupling comes into play. If we instead follow Eq. (6.11) of \cite{Fairbairn:2019xog}, we do not need to modify our calculations, apart from the replacement of $\alpha_{tr}$ by $\alpha_d$.)} 
Similar to the prescription employed to evaluate the Chapman-Jouguet speed, the efficiency factor $\kappa_s$, which is generally a function of the bubble wall velocity and phase transition strength, must be evaluated at $\alpha_d$ since the FOPT occurred in the dark sector \cite{Fairbairn:2019xog,Nakai:2020oit}. Assuming that the FOPT falls in the nonrunaway regime, we have
    \begin{eqnarray}
        \kappa_s\left(\alpha_{d}\right) \simeq \begin{cases}
            \frac{\alpha_{d}}{0.73+0.083\sqrt{\alpha_{d}}+\alpha_{d}},~v_w \geq v_{w,\alpha}\\
            \frac{6.9\alpha_{d} v_w^{6/5}}{1.36-0.037\sqrt{\alpha_{d}}+\alpha_{d}},~v_w \leq v_{w,\alpha}
        \end{cases},
    \end{eqnarray}
    where
    \begin{eqnarray}
        v_{w,\alpha}\left(\alpha_d\right) \equiv \left[\frac{1.36-0.037\sqrt{\alpha_{d}}+\alpha_{d}}{6.9\left(0.73+0.083\sqrt{\alpha_{d}}+\alpha_{d}\right)}\right]^{5/6}.
    \end{eqnarray}
One of the usual methods employed to determine the sensitivity of GW experiments to GW signals produced, \textit{e.g.} through a FOPT, is to check if the predicted GW abundance $\Omega_{sig}(f) h^2$ is above the GW sensitivity curve for a given experiment $\Omega_{noise}(f) h^2$, for some frequency $f$ within the operating frequency band $[f_{min},f_{max}]$ of the experiment. Another method requires that the SNR, defined as
\begin{eqnarray}
\label{SNRGWdef}\rho^2 \equiv n_{det} \tau_{obs,GW} \int_{f_{min}}^{f_{max}} df~\left[\frac{\Omega_{sig}(f) h^2}{\Omega_{noise}(f) h^2}\right]^2,
\end{eqnarray}
is above some threshold value $\rho_{th}$. Here, $\tau_{obs,GW}$ is the expected runtime of the experiment, and $n_{det} = 1$ for an autocorrelation measurement and $n_{det}=2$ for a cross-correlation measurement (\textit{e.g.} aLIGO, aVirgo, KAGRA, CE, ET, LISA, DECIGO, BBO, PTA). However, reference \cite{Schmitz:2020syl} points out disadvantages of these two methods, and instead advocates the idea of introducing \textit{peak-integrated sensitivity curves} (PISCs). The main assumption of this framework is the factorizability of the GW abundance spectrum into a peak value---a model dependent quantity---multiplied by a spectral function, whose form is model independent. Then we can write
\begin{eqnarray}
\label{Factorization}\Omega_{sig}(f)h^2 = \Omega_{peak} h^2~\mathcal{S}(f,f_s).
\end{eqnarray}
As an illustration, one can take the case of GW produced through sound waves; the peak GW abundance is given by Eq. (\ref{Omh2PeakSound}) while the spectral function $\mathcal{S}$ is given by Eq. (\ref{SigSpexSound}). The ansatz given in Eq. (\ref{Factorization}) allows us to write
\begin{eqnarray}
\label{SNRsimple}\rho(f_s) = \frac{\Omega_{peak} h^2}{\Omega_{PIS}(f_s) h^2} ,
\end{eqnarray}
where
\begin{eqnarray}
\label{PISdef}\left(\Omega_{PIS}h^2\right)^{-2}(f_s) \equiv n_{det}\tau_{obs,GW}\int_{f_{min}}^{f_{max}}df~\left[\frac{\mathcal{S}(f,f_s)}{\Omega_{noise}(f)h^2}\right]^2.
\end{eqnarray}
As defined in Eq. (\ref{PISdef}), the PIS only depends on the shape of the noise spectrum intrinsic to the GW experiment or mission, and on the form of the spectral function. Once we have the PIS for an experiment/mission and for a specific source of GW signal, the SNR can immediately be obtained from Eq. (\ref{SNRsimple}) once we have the model dependent quantities, namely the peak GW signal and the corresponding peak frequency, without repeating the integration Eq. (\ref{SNRGWdef}) for each set of model parameters.

\begin{figure}[t!]
    \centering
    \includegraphics[scale=0.36]{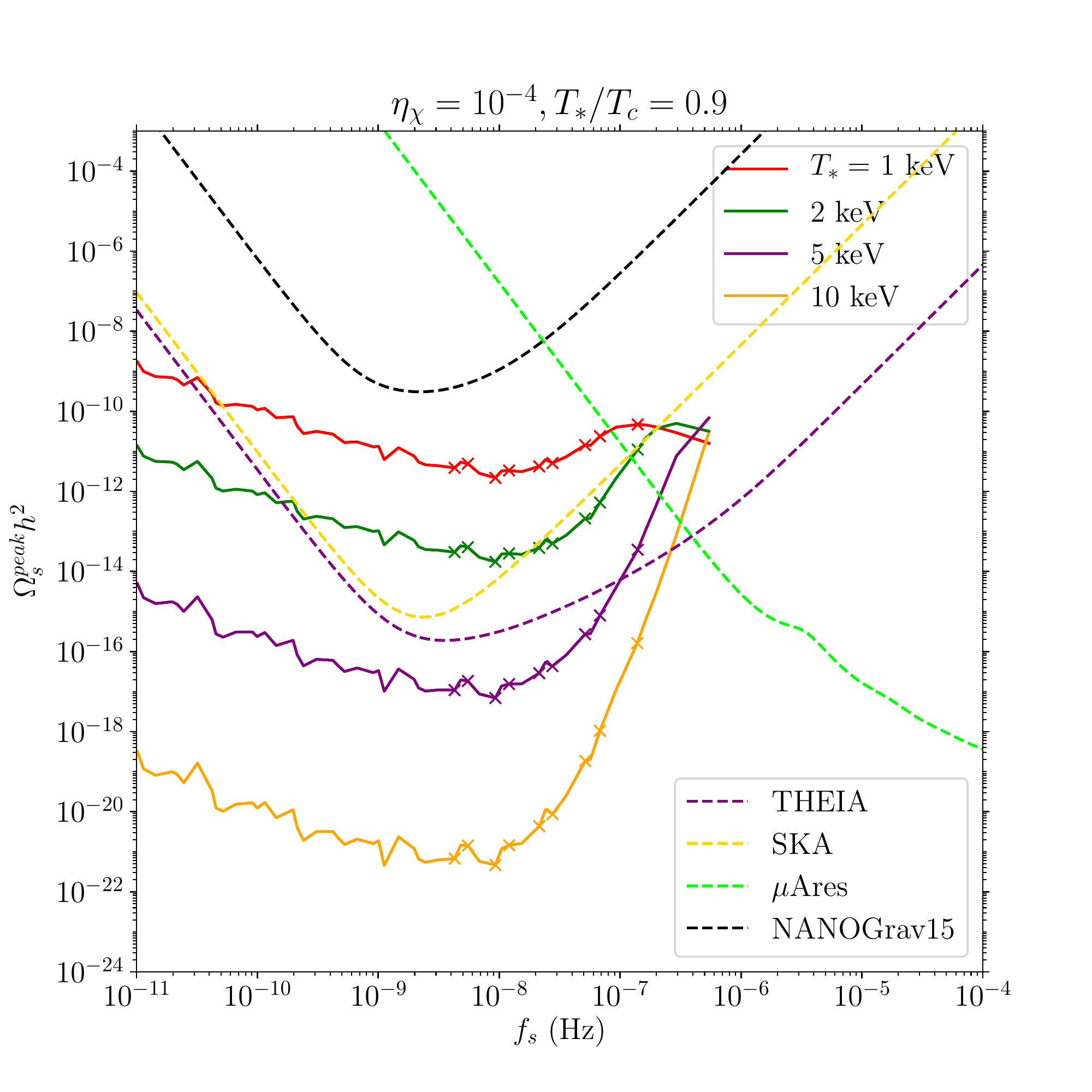}
    \includegraphics[scale=0.36]{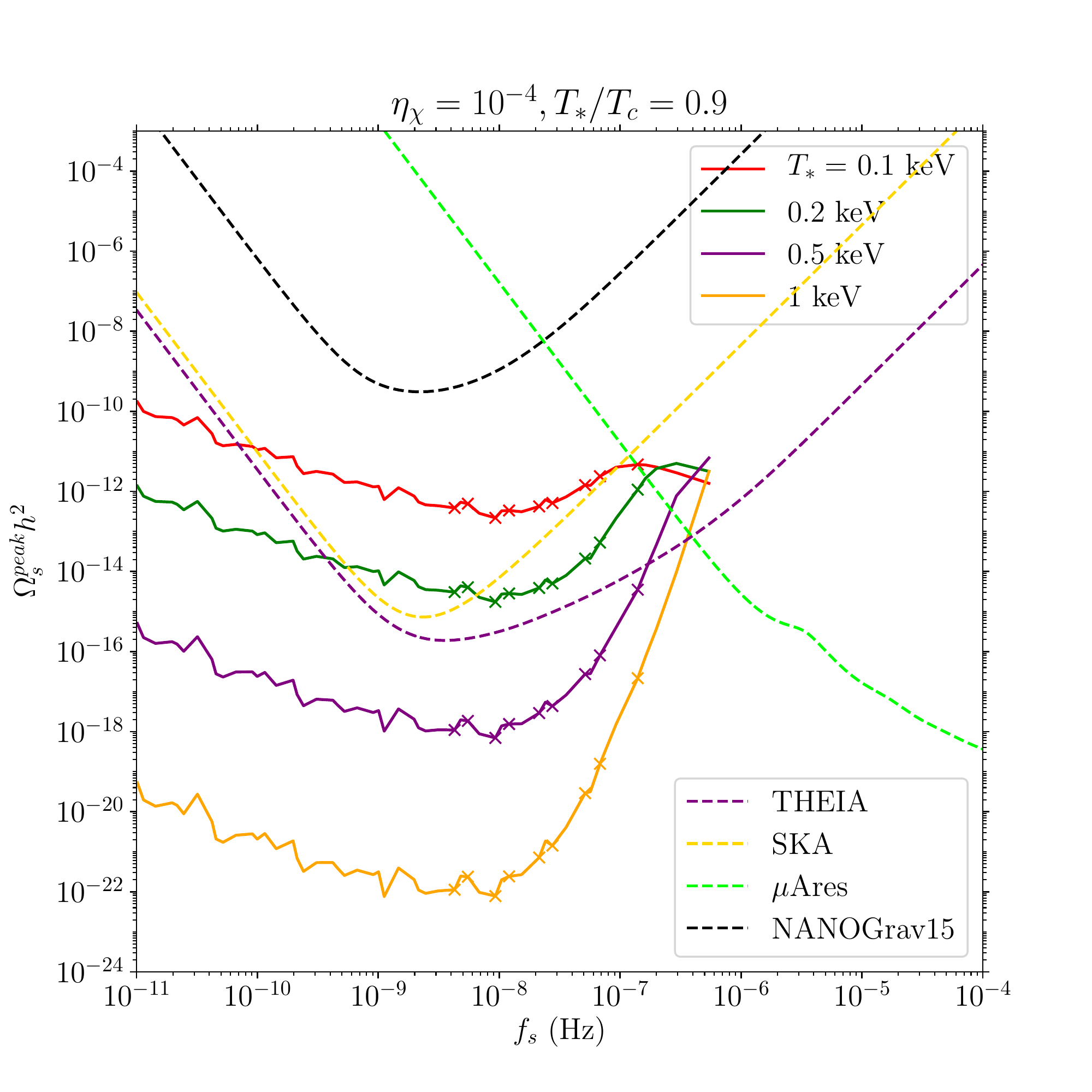}\\
    \includegraphics[scale=0.36]{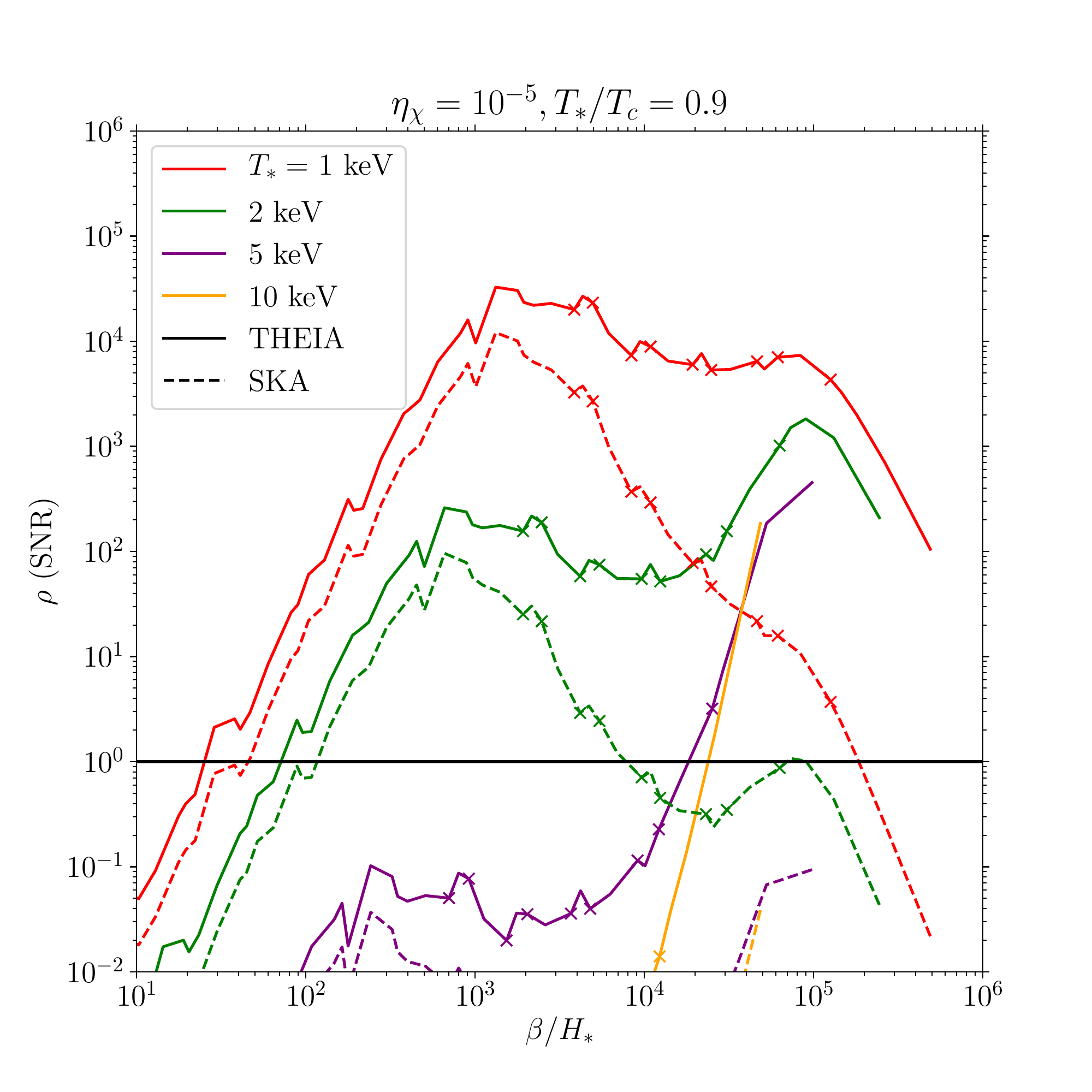}
    \includegraphics[scale=0.36]{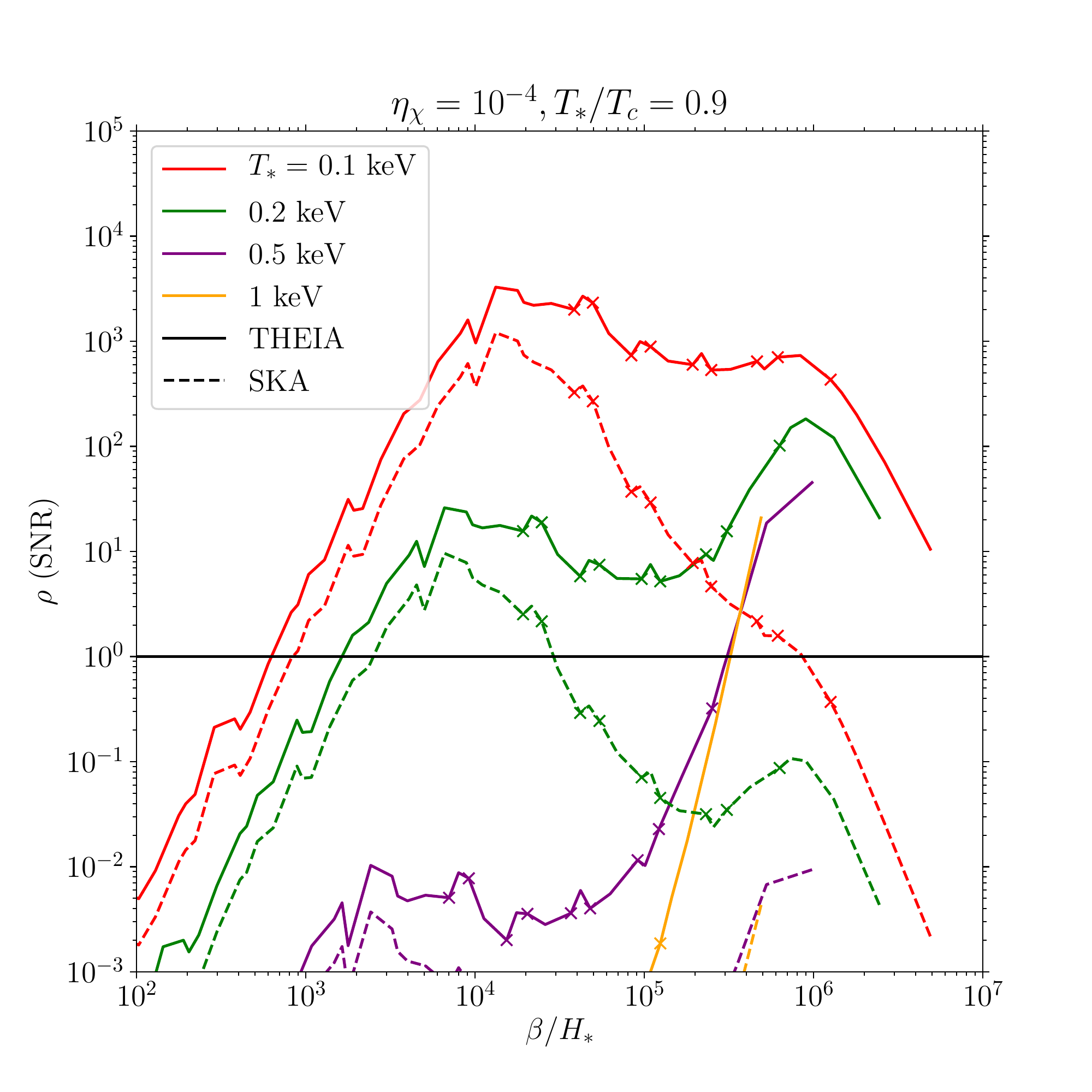}
    \caption{\label{fig:GW}In both top row panels, the solid lines are projected curves on the GW abundance-frequency plane of points from Fig. \ref{fig:PBHFOPTSens}, corresponding to PBH formation scenarios that can be probed by an SKA-like PTA observation, for the benchmark cases where $\eta_\chi = 10^{-5}$ (left) and $\eta_\chi = 10^{-4}$ (right), and $T_*/T_c = 0.9$. The dashed curves are the computed PISCs for THEIA, SKA, $\mu$Ares, and NANOGrav (15-yr data) strain sensitivities, using the prescription in \cite{Schmitz:2020syl}, assuming that: (i) the model independent part of the signal spectrum corresponds to a GW produced from sound waves; and (ii) the observation period is $\tau_{obs,GW} = \unit[20]{yrs}$.  As in Fig. \ref{fig:PBHFOPTSens}, the cross marks correspond to average PBH mass values  $10^{-5}, 5 \times 10^{-6}, 10^{-6}, 5 \times 10^{-7}, 10^{-7}, 5 \times 10^{-8}, 10^{-8}, 5 \times 10^{-9}$, and $10^{-9} M_\odot$. For the bottom row panels, we plot the GW signal SNR for THEIA (solid) and SKA (dashed) as a function of $\beta/H_*$, for the same set of benchmark FOPT parameters.}
\end{figure}

In our scenario where GWs are produced from sound waves during a FOPT in the dark sector, specifying the FOPT parameters will provide a prediction for the peak GW abundance and the corresponding frequency, as provided in Eqs. (\ref{Omh2PeakSound}) and (\ref{PeakFreqSound}), respectively. In Fig. \ref{fig:GW}, we projected the FOPT parameters from Fig. \ref{fig:PBHFOPTSens}, for $T_*/T_c = 0.9$, and $\eta_\chi = 10^{-4}$ and $\eta_\chi = 10^{-5}$, corresponding to PBH formation scenarios that can be probed by PTA, on the $\Omega_s^{peak} h^2-f_s$ plane. The cross marks indicate the same set of average PBH mass values as in Fig. \ref{fig:PBHFOPTSens}, where $\langle M_{PBH}\rangle$ increases with $f_s$. On the other hand, the PISCs corresponding to THEIA, SKA, and $\mu$Ares are shown as dashed lines. Note that $\mu$Ares is a proposed space-based GW antenna \cite{Sesana:2019vho} that will be expected to operate in the $\mu$Hz frequency band, with better sensitivity compared with LISA. Meanwhile, THEIA \cite{Theia:2017xtk} is another proposed mission designed to perform high precision astrometry at an accuracy level that is 1000 times better than the ESA Gaia mission. The principle behind detecting stochastic GW signals using astrometry is based on the idea that such GW signals will produce correlated angular displacements of stars, and monitoring of the proper motions of a collection of stars can be used to place limits on the GW abundance \cite{Book:2010pf,Garcia-Bellido:2021zgu}. As for SKA \cite{Janssen:2014dka}, gravitational wave astronomy can be performed in an analogous manner, as laid down in, \textit{e.g.} \cite{HellingsDowns}, by looking at the effect of GWs on the timing signal of a collection of pulsars, and the correlation between the changes in the timing signals of the observed pulsars can be used to infer the strength of the GW signal. For all PISCs, we assume a 20-year observation period; for any other choice for the observation period, the PISC must be rescaled by an appropriate factor that is proportional to $\tau_{obs,GW}^{1/2}$. Conversely, a prediction for the GW SNR that is a factor of 10 larger than the SNR threshold for a GW experiment, for a given observation time $\tau_{obs}$, means that the required observation time to reach the SNR threshold is reduced by a factor of 100.

Looking at the PISCs in the top row of Fig. \ref{fig:GW}, THEIA provides the most stringent constraints on stochastic GW signals in the peak frequency range around $\unit[10^{-10}-10^{-7}]{Hz}$, with SKA limits being competitive in the subnanohertz range. Beyond the microhertz range, both SKA and THEIA lose sensitivity, while $\mu$Ares provides better reach to GW abundances as low as $\sim 10^{-18}$, for peak frequencies of at most $\unit[10^{-4}]{Hz}$. For the FOPT parameters that can be probed by pulsar timing, the most relevant constraints come from THEIA and $\mu$Ares; meanwhile, since the SKA PISC is enveloped by the THEIA PISC, THEIA will be sensitive to any FOPT scenario that yields a GW signal SNR for SKA above unity. For the projected FOPT parameter points, where $T_*/T_c$ is fixed to 0.9, the corresponding SNR can be directly read off from Fig. \ref{fig:GW} using Eq. (\ref{SNRsimple}). In the case of $\eta_\chi = 10^{-5}$, we found that most of the parameter points along the PTA sensitivity curve, corresponding to percolation temperatures of \unit[1-2]{keV}, can produce GWs that can be probed by SKA, assuming that the threshold SNR is set to 1; for percolation temperatures of at least \unit[5]{keV}, SKA is not sensitive to the associated GW signal. Meanwhile, for $\eta_\chi = 10^{-4}$, SKA is sensitive to FOPT scenarios with $T_* = \unit[0.1-0.2]{keV}$, and the SNR goes below threshold for higher percolation temperatures, as shown in the top right panel of Fig. \ref{fig:GW}. This trend in the SNR, attributed to the decrease in the peak GW abundance in the frequency range below $\sim\unit[10^{-8}]{Hz}$, can be understood by looking at Eq. (\ref{Omh2PeakSound}): the behavior of the peak GW abundance $\Omega_s^{peak} h^2$ is mainly driven by the decrease in $\alpha_{tr}$, which can happen as the result of increasing percolation temperature, according to the $\alpha_{tr}$ estimate in Eq. (\ref{alph_est}). 

In light of the recent discovery of low frequency gravitational waves by NANOGrav based on 15-year observational data \cite{NANOGrav:2023gor}, one can check that that NANOGrav is quite insensitive to the GWs produced in our mechanism, assuming the parameter range of interest. Given the strain sensitivity $h_c(f)$ provided in \cite{NANOGrav:2023ctt}, the corresponding GW abundance can be computed via, \textit{e.g.} \cite{Garcia-Bellido:2021zgu,Yokoyama:2021hsa},
\begin{eqnarray}
    \Omega_{GW}h^2 = \frac{2\pi^2 h^2}{3H_0^2}f^2 h_c^2(f).
\end{eqnarray}
We then calculated the PISC and display the resulting curve in Fig. \ref{fig:GW}, and one can see that the recent NANOGrav result has little to no direct impact on our setup.

Focusing on the proposed THEIA experiment, the bottom row panels of Fig. \ref{fig:GW} show the corresponding GW SNR for the same set of benchmark FOPT scenarios. We generally find a sharp change in sensitivity for $\beta/H_*$ in the \unit[2-5]{keV} range for $\eta_\chi = 10^{-5}$, and in the \unit[0.2-0.5]{keV} range for $\eta_\chi = 10^{-4}$. In this range of percolation temperatures, the peak frequency where the projected PTA sensitivity curve crosses the THEIA PISC is below $\sim$nHz where the THEIA PISC is decreasing. Beyond a certain percolation temperature, the peak frequency is above nanohertz where the THEIA PISC is increasing; this can only happen if $\alpha_{tr}$ increases with the percolation temperature. From Eq. (\ref{alph_est}), one can increase the PBH fraction, which corresponds to lighter PBH masses in the PTA sensitivity curve. In turn, lighter PBH masses can be achieved if the FV bubble, which trapped the DM particles that formed the FB progenitor, has a smaller radius: this amounts to a fairly large dimensionless FOPT rate $\beta/H_*$. This is also the reason why the projected PTA sensitivity curves that lie above the THEIA PISC correspond to the low $\langle M_{PBH}\rangle$ tail, for higher percolation temperatures. 

It is then worth emphasizing that for benchmark FOPT scenarios with DM asymmetry $\eta_\chi = 10^{-5}$ ($10^{-4}$), $T_*$ at \unit[2]{keV} ($\unit[0.2]{keV}$) or below, and $\beta/H_*$ around the  $\mathcal{O}(10^2)-\mathcal{O}(10^4)$ range, these produce PBHs that can be probed through an SKA-like PTA observation of Doppler phase shifts; on the other hand, using SKA to look for GW signals from a FOPT will also be sensitive to these FOPT scenarios. This may provide a strong motivation for including the search for Doppler shifts in the pulsar timing signal as one of the goals of SKA, whether or not a stochastic GW signal can be detected by SKA, albeit a lack of a GW signal and a presence of a PTA signal would be inconclusive in validating the PBH formation scenario discussed in this study. It might also be possible that a GW signal will be found, but not a PTA signal: this may occur for FOPT scenarios with high percolation temperatures, which lead to light PBHs beyond the reach of pulsar timing. Certainly, a simultaneous detection of stochastic GW signals and Doppler shift in the phase signal would be an ideal situation in probing this FOPT scenario.
\section{Generic quartic potential}
\label{sec:potential}
At this point, we were able to identify the FOPT parameters, that produce PBHs that can be probed by PTA, and also generate GWs that can potentially be detected through correlated shifts in positions of a collection of stellar objects, or through a space-based GW antenna. In Sec. \ref{sec:PBHFOPT}, we have introduced the effective potential $V_{eff}$ for the dark sector scalar field $\phi$, from which all relevant quantities that describe the FOPT originate. The goal now is to obtain the class of models which lead to the desired FOPT parameters that can be probed by PTA and stochastic GW observations. We can take a generic form for the effective potential $V_{eff}$ following \cite{Marfatia:2021hcp}, so that
\begin{eqnarray}
\label{VeffOrig}V_{eff}(\phi,T) = D\left(T^2 - T_0^2\right)\phi^2-\left(AT+C\right)\phi^3+\frac{\lambda}{4}\phi^4. 
\end{eqnarray}
The parameter $T_0$, the destabilization temperature, can be traded with the energy gap between the false and true vacua at zero temperature, which we shall denote as $B$. One can show that the true vacuum configuration $\phi_0$ at zero temperature satisfies the following equations
\begin{eqnarray}
-B &=& -D T_0^2 \phi_0^2 - C\phi_0^3 + \frac{\lambda}{4}\phi_0^4\\
0 &=& -2DT_0^2 - 3C \phi_0 + \lambda \phi_0^2,
\end{eqnarray}
so that we have
\begin{eqnarray}
\label{T0eqn}T_0^2 &=& \frac{9C^2}{8D\lambda}\left[\zeta^2(B) - 1\right],
\end{eqnarray}
where $\zeta(B)$ is the positive root to the polynomial equation
\begin{eqnarray}
\label{quarticequation}\zeta^4 + \frac{8}{3}\zeta^3 + 2\zeta^2 - \left(\frac{1}{3}+\frac{4B}{\lambda K^4}\right)=0, \quad K \equiv \frac{3C}{2\lambda}.
\end{eqnarray}
From the effective potential, we can also extract the critical temperature $T_c$ at which the true and false vacua are degenerate in energy; we find that $T_c$ is the positive root of the equation
\begin{eqnarray}
\label{Tceqn}\left(A^2 - \lambda D\right)T_c^2 + 2AC T_c + \left(C^2 + \lambda D T_0^2\right)=0.
\end{eqnarray}
Taking the fundamental parameters that characterize the effective potential to be 
\begin{eqnarray}
    \nonumber \{B^{1/4}, C, D, \lambda\},
\end{eqnarray}
we can identify the physical conditions that guarantee the existence of $T_0$ and $T_c$. Requiring that we obtain positive roots from Eqs. (\ref{quarticequation}) and (\ref{Tceqn}), we have
\begin{eqnarray}
    \label{PhysicalCondition}B^{1/4} > \left(\frac{4\lambda}{3}\right)^{1/4}K,\quad A^2 - \lambda D < \left(\frac{C}{T_0}\right)^2,\quad K \equiv \frac{3C}{2\lambda}.
\end{eqnarray}
From the expression for the effective potential, Eq. (\ref{VeffOrig}), the VEV, energy gap between false and true vacua, and both their derivatives, are given by
\begin{eqnarray}
\phi_+(T) &=& \frac{3(AT+C)+\sqrt{9(AT+C)^2-8D\lambda\left(T^2-T_0^2\right)}}{2\lambda},\quad T < T_c,\\
\Delta V_{eff} &=& \frac{1}{3}\left[\frac{\lambda}{4}\phi_+^4(T)-D\left(T^2-T_0^2\right)\phi_+^2(T)\right],\\
\nonumber T\frac{d}{d T}\Delta V_{eff} &=& \frac{1}{3}\left\{-2DT^2\phi_+^2(T)+T\frac{d\phi_+}{dT}\left[\lambda\phi_+^3(T)-2D\left(T^2-T_0^2\right)\phi_+(T)\right]\right\},\quad T < T_c,\\
~\\
\frac{d\phi_+(T)}{dT} &=& \frac{3A\phi_+(T)-4DT}{2\lambda \phi_+(T)-3(AT+C)},\quad T < T_c.
\end{eqnarray}
The above expressions can be readily used to calculate the FOPT strength from Eqs. (\ref{TraceAnomaly}) and (\ref{AlphaTr}). Furthermore, obtaining the FOPT rate $\beta$ requires knowledge of the nucleation rate provided in Eq. (\ref{S3NucRate}). For a certain class of quartic effective potentials for a single scalar field, \cite{Adams:1993zs} obtained a semianalytic expression for the O(3)-symmetric bounce action given by
\begin{eqnarray}
S_3(T) &=& \frac{\pi a}{\bar{\lambda}^{3/2}}\frac{8\sqrt{2}}{81}\left(2-\delta\right)^{-2}\sqrt{\frac{\delta}{2}}\left(\beta_1 \delta +\beta_2 \delta^2+\beta_3 \delta^3\right),\\
\beta_1 = 8.2938,\quad &\beta_2& = -5.5330,\quad \beta_3 = 0.8180,
\end{eqnarray}
where
\begin{eqnarray}
\delta \equiv \frac{8\bar{\lambda} b}{a^2},\quad \bar{\lambda} \equiv \frac{\lambda}{4},\quad a(T) \equiv AT+C,\quad b(T) \equiv D\left(T^2 - T_0^2\right),\quad D > 0.
\end{eqnarray}
Such quartic effective potentials are assumed to have two local minima, where each minimum can be associated with either the true/false vacuum, separated by a barrier. We require that $\lambda > 0$ so that the potential is always bounded from below. Note that $0 < \delta < 2$, where $\delta \simeq 2$ corresponds to the thin wall limit \cite{Coleman:1977py}, where the two potential minima are nearly degenerate in energy. The FOPT rate $\beta$ at the percolation temperature $T_*$ is then given by
\begin{eqnarray}
\label{betaVeff}\frac{\beta}{H_*} \simeq T_*\frac{d}{dT}\left[\frac{S_3}{T}\right]\Bigg\vert_{T_*}.
\end{eqnarray}
The time-temperature relation is obtained from entropy conservation in the SM and the dark sector, assuming that they are only coupled through gravity. Since the duration of the phase transition is rather short relative to the Hubble expansion, and assuming that we are away from mass thresholds, we can assume that the effective number of relativistic degrees of freedom is constant in either sector. Furthermore, the dark and visible sectors do not exchange energy and entropy, so the entropy in each sector is separately conserved. This implies that the temperature ratio in the two sectors is constant, and this condition was invoked in Sec. \ref{sec:PBHFOPT}. Noting that 
\begin{eqnarray}
H \simeq H_c \left(\frac{T_{SM}}{T_{SM,c}}\right)^2,\quad H_c \equiv H(T_{SM}=T_c/\xi,T_c),
\end{eqnarray}
one can show that
\begin{eqnarray}
2H_c t(T_{SM}) &\simeq& 1+\left(\frac{T_{SM,c}}{T_{SM}}\right)^2,\\
-\ln f_{FV} &\simeq& \frac{4\pi}{3}v_w^3~\frac{1}{8H_c^4}\int_y^1 \frac{dy'}{y'^3}\left(\frac{1}{y^2}-\frac{1}{y'^2}\right)^3\Gamma(T_c y'),\quad y \equiv \frac{T_{SM}}{T_{SM,c}}.
\end{eqnarray}
\begin{figure}
    \centering
    \includegraphics[scale=0.37]{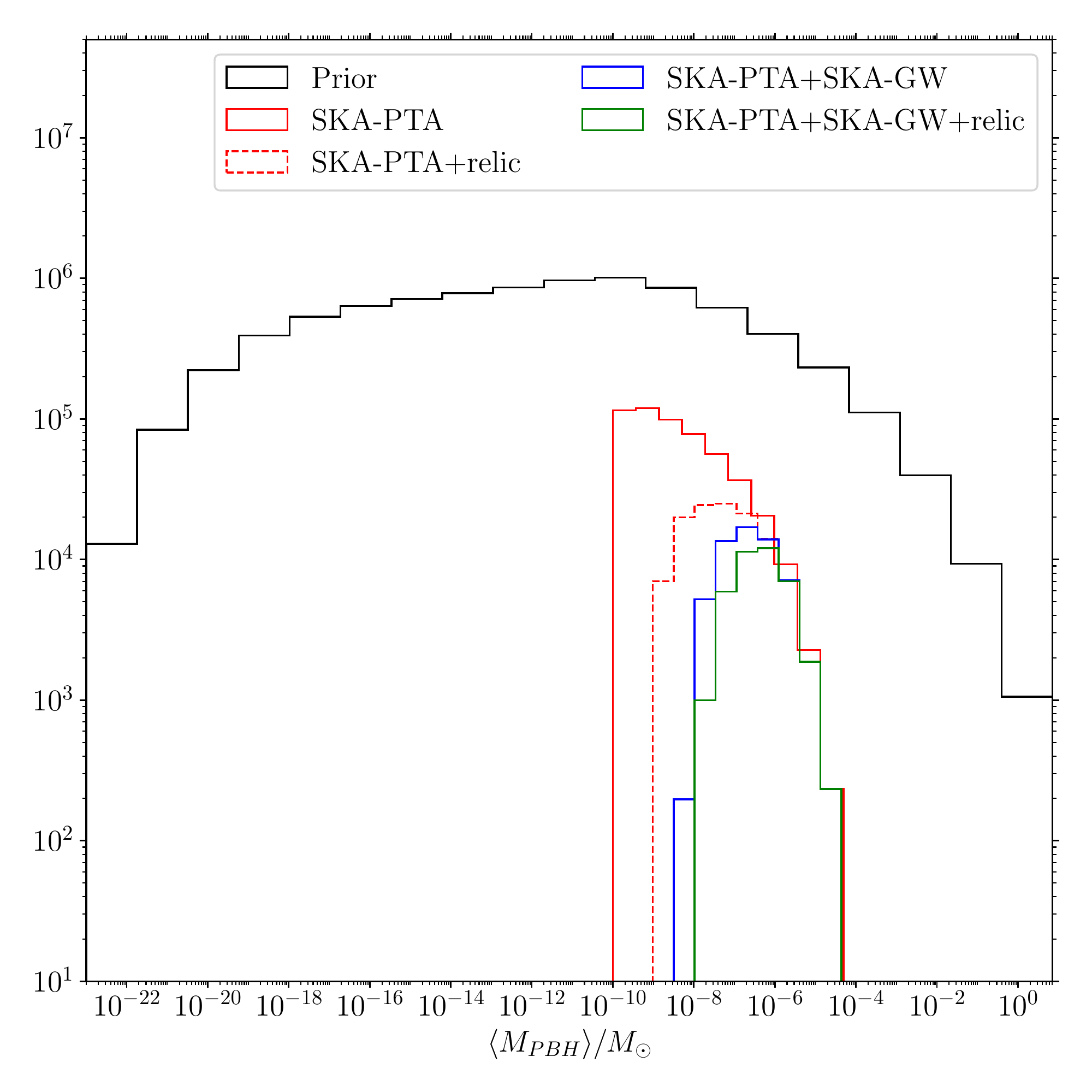}\includegraphics[scale=0.37]{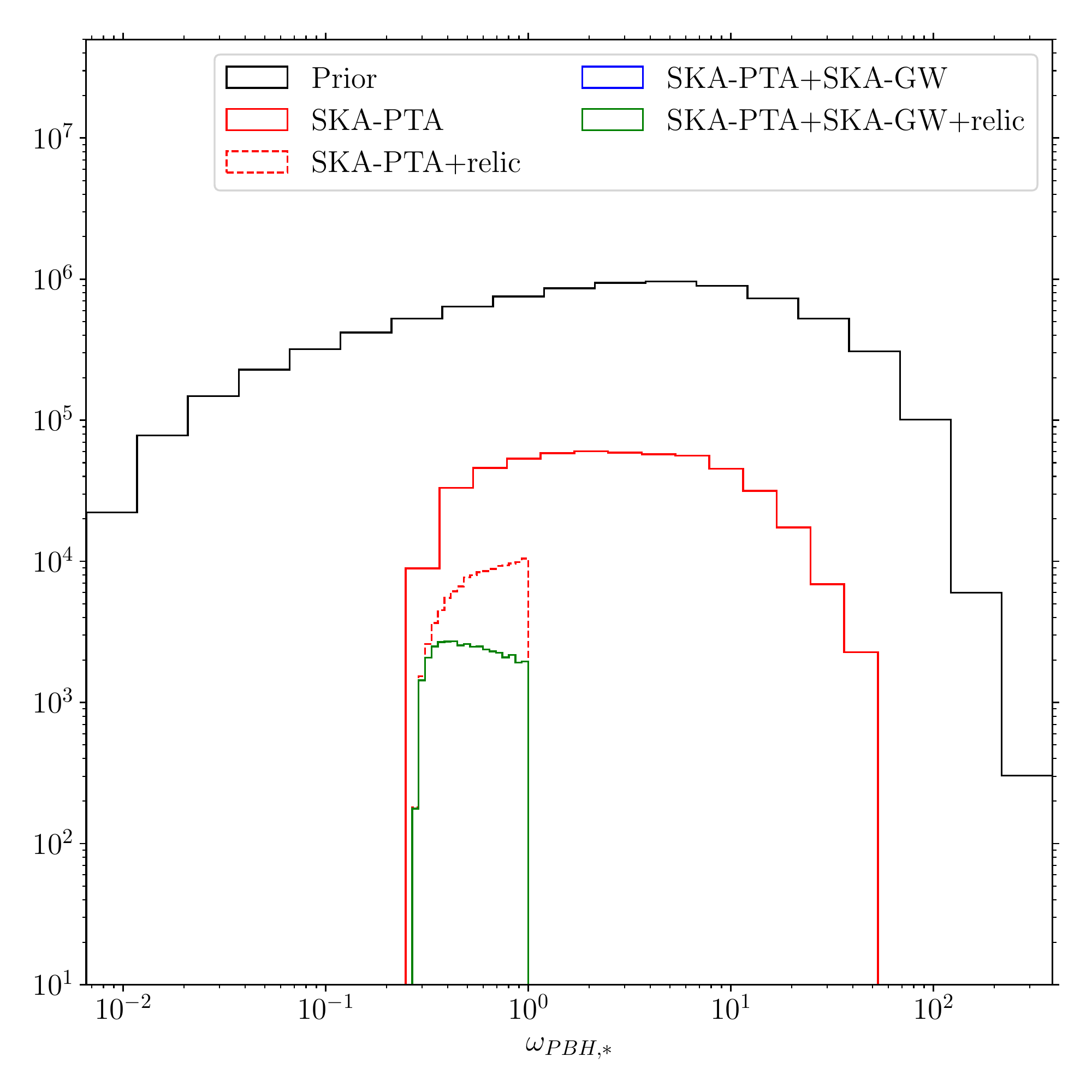}\\
    \includegraphics[scale=0.37]{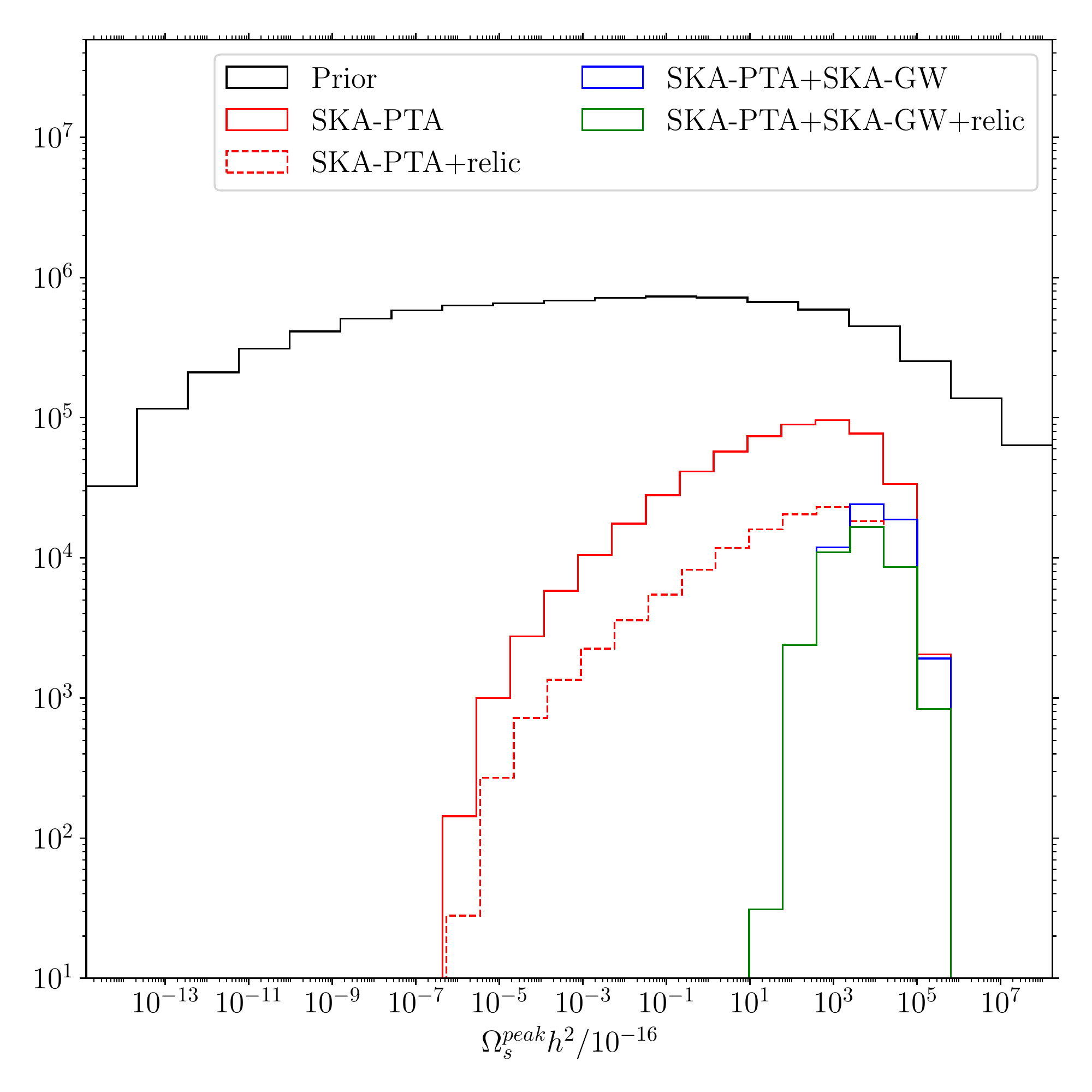}\includegraphics[scale=0.37]{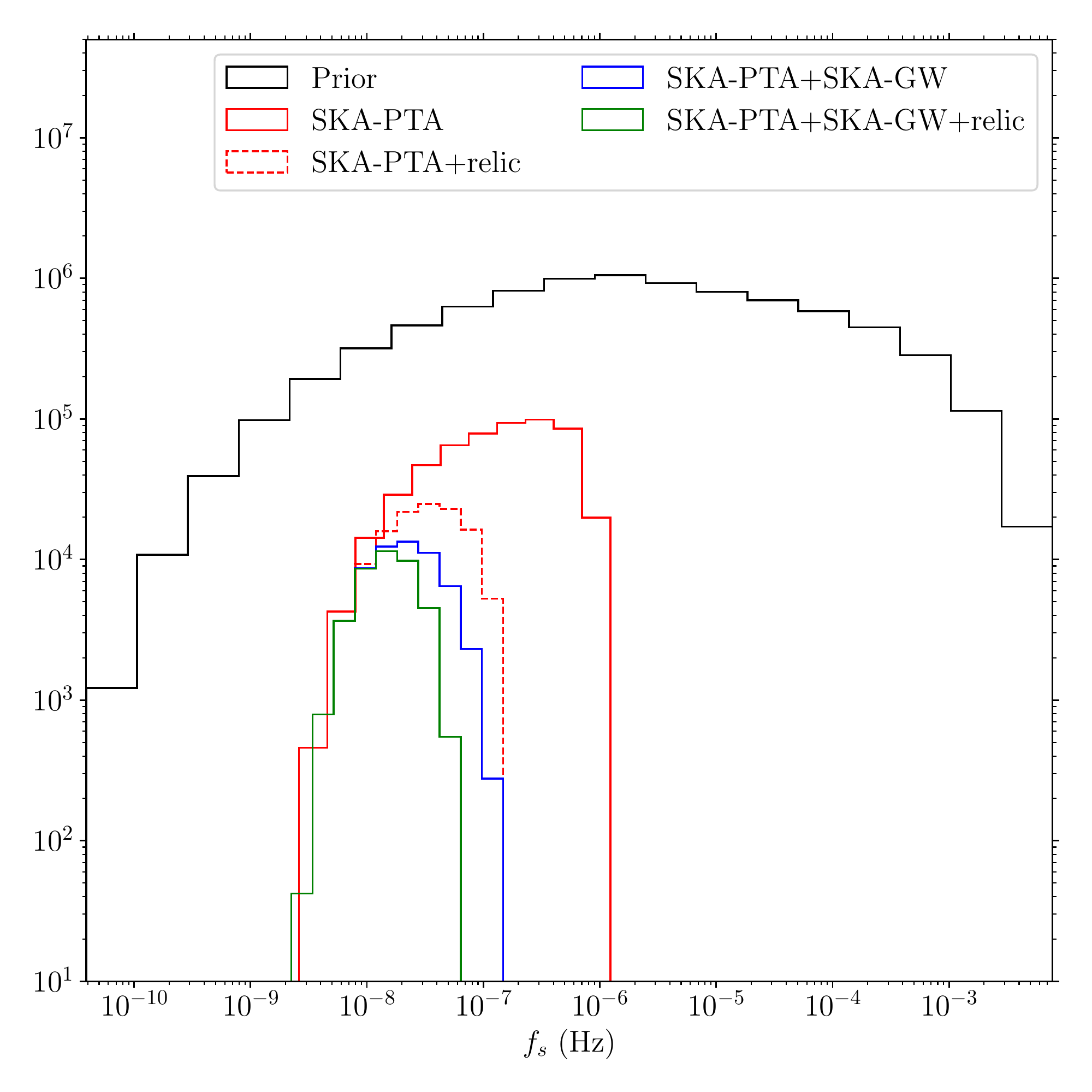}
    \caption{\label{fig:HistPhys} Plots of the one dimensional histograms for the physical parameters $\langle M_{PBH}\rangle$, $\omega_{PBH,*}$, $\Omega_s^{peak} h^2$, and $f_s$. The black lines indicate the initial 1D distribution of the physical parameters, labelled as ``Prior," given a uniform distribution of points in the hypercube where we performed the scan on the fundamental parameters in the effective potential. The solid red histograms refer to the distribution of parameter points that yield a detectable Doppler shift in an SKA-like pulsar timing observation, while the dashed red histograms include a subset of points that yield PBH fractions less than unity. The blue histograms refer to FOPT scenarios that can be probed in SKA, both by looking at Doppler shifts in the pulsar timing signal, and stochastic GW signals; additional constraint from the PBH relic density being less than that of DM is represented by the green histograms.}
\end{figure}
\begin{figure}
    \centering
    \includegraphics[scale=0.49]{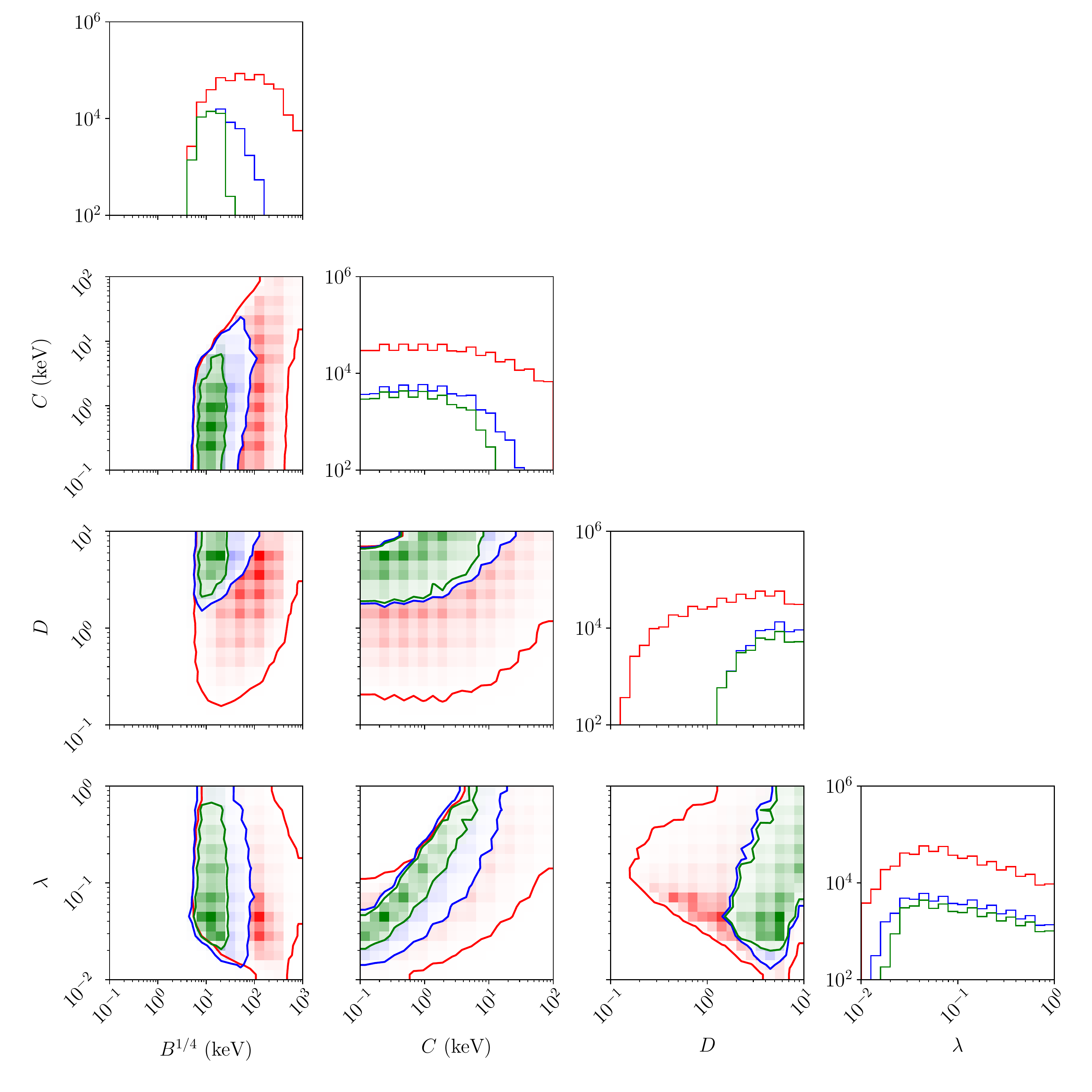}
    \caption{\label{fig:VeffScan} We present the result of the scan over the parameters $\{B^{1/4}, C, D, \lambda\}$ appearing in the effective potential. For the two dimensional corner plots, the red curves bound the parameter regions which lead to FOPT scenarios where the produced PBHs are detectable through Doppler signal from an SKA-like pulsar timing measurement. The blue curves enclose those points in which GWs in the form of sound waves, generated during the FOPT, can be probed by SKA; the green curves enclose those points that also lead to PBH fractions less than unity. Those points lying outside these curves are either unphysical points, or physical points that cannot be probed via pulsar timing nor GW observations.}
\end{figure}
\begin{figure}
    \centering
    \includegraphics[scale=0.47]{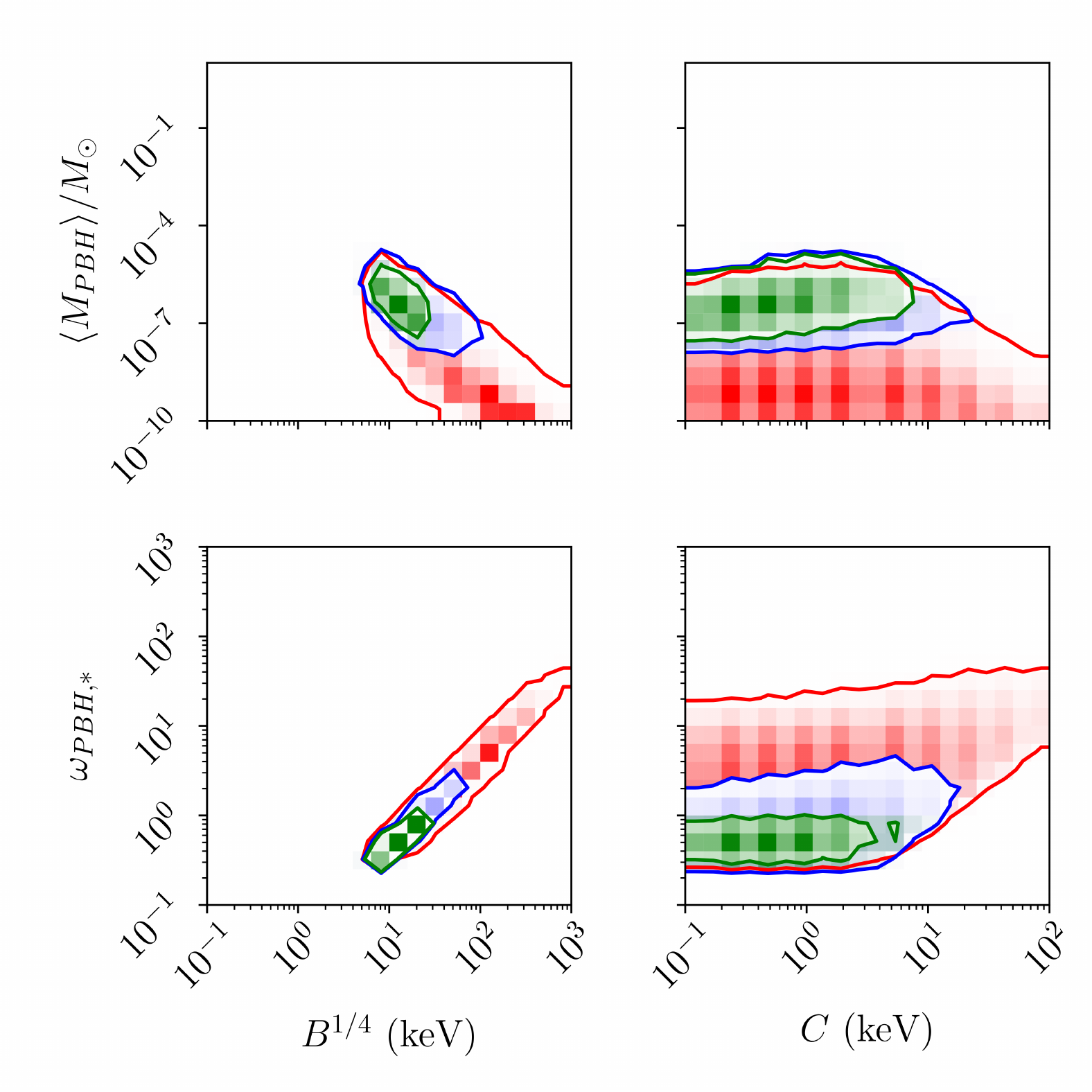}\includegraphics[scale=0.47]{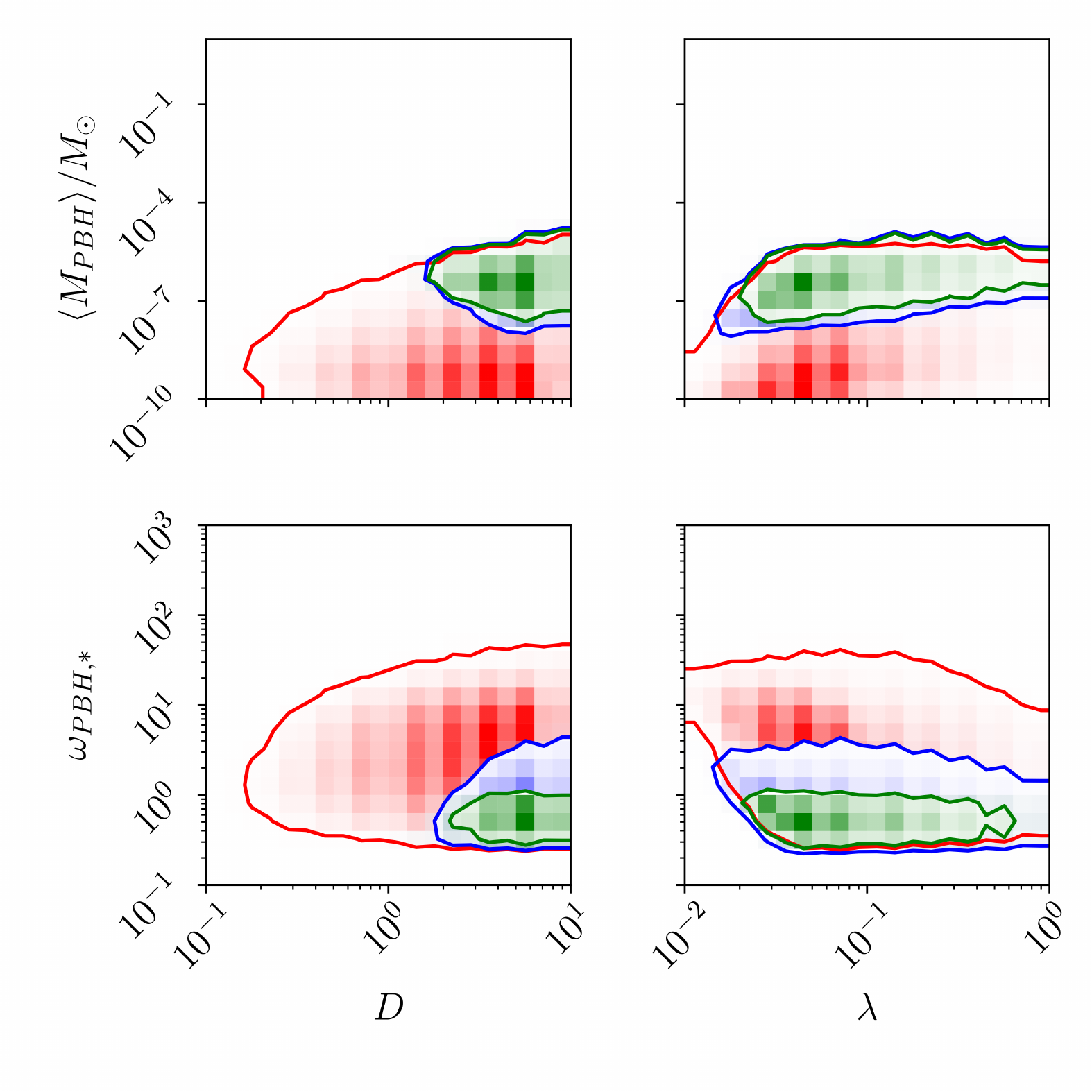}\\
    \includegraphics[scale=0.47]{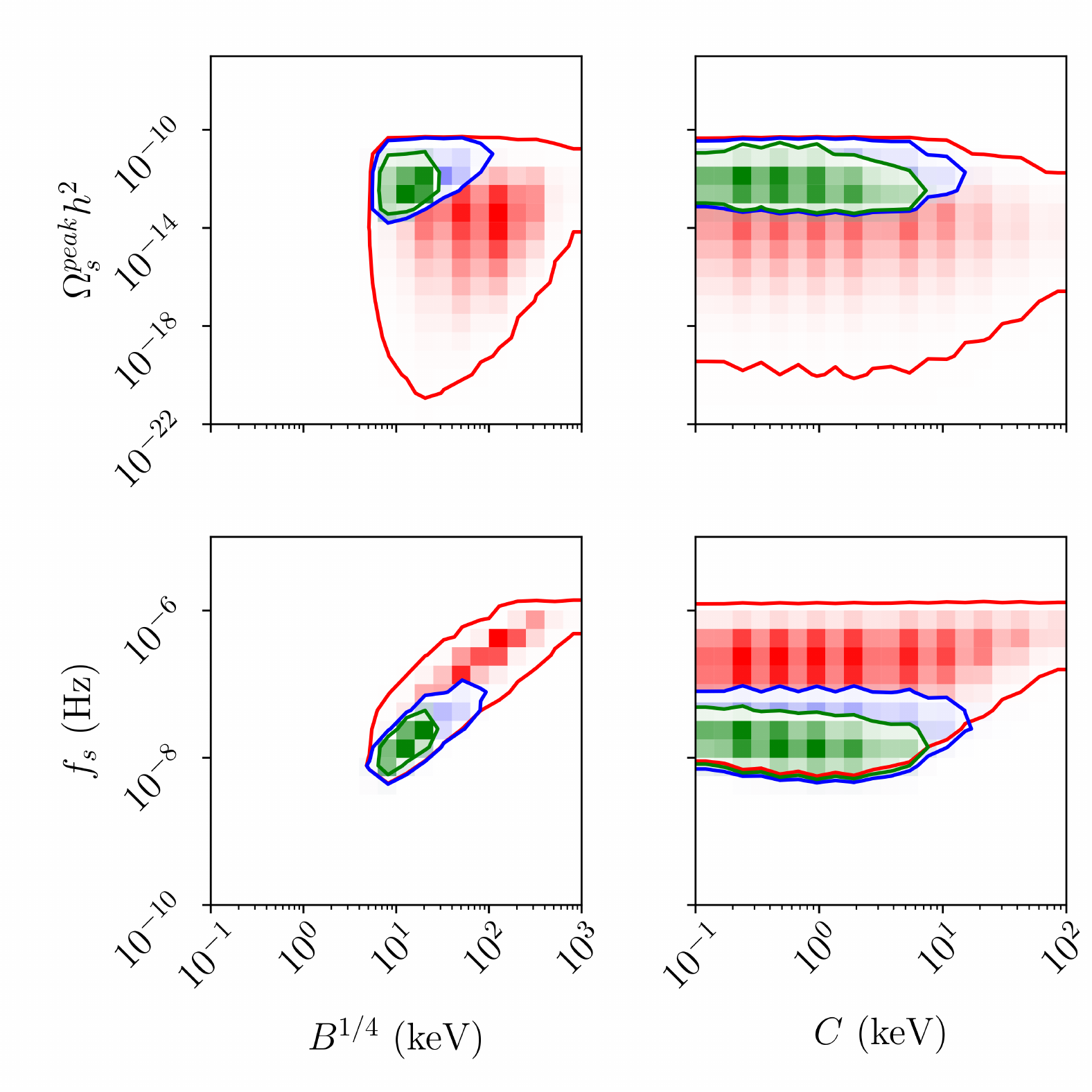}\includegraphics[scale=0.47]{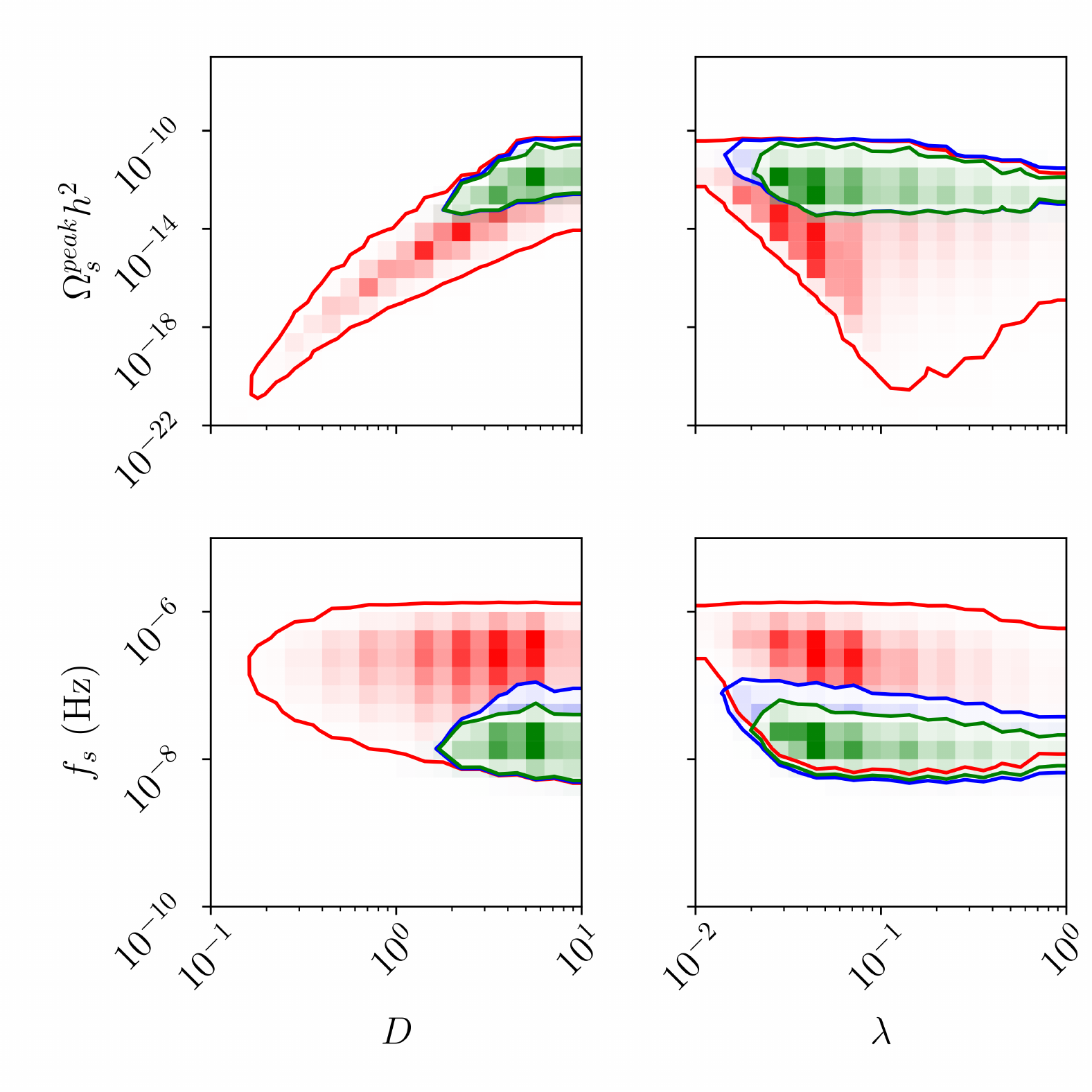}
    \caption{\label{fig:PhysCorrScan} Similar to Fig. \ref{fig:VeffScan}, we show the correlations of the fundamental parameters in the effective potential with the physical parameters that can be probed in PBH searches via Doppler shifts in pulsar timing signals and GW searches.}
\end{figure}
Fixing $A = 0.1$ and $\eta_\chi = 10^{-5}$, we performed the scan over the parameters $\lbrace B^{1/4}, C, D, \lambda\rbrace$.  We restricted our scan within
\begin{eqnarray}
    \nonumber \unit[10^{-1}]{keV} \leq B^{1/4} \leq \unit[10^3]{keV}&,&\quad \unit[10^{-1}]{keV}\leq C \leq \unit[10^2]{keV},\\
    \nonumber 10^{-1} \leq D \leq 10&,&\quad 10^{-2} \leq \lambda \leq 1.
\end{eqnarray}
An initial summary of the results of the scan is shown in Fig. \ref{fig:HistPhys}, where we plot one dimensional histograms of the physical observables in SKA, namely: the average PBH mass and PBH fraction; and the peak GW abundance and peak GW frequency. Starting from flat priors for the fundamental parameters appearing in the effective potential, the solid black histograms represent the resulting distributions of the physical observables. We note that we only included parameter points in the regime where 
$\vert T_* - T_c\vert < 0.1 T_c$ and $\alpha_{tr}$ is sufficiently small such that the dark plasma will not be reheated back to the critical temperature. The scanned hypervolume leads to a fairly wide range of: average PBH masses, from $\sim 10^{-22} M_\odot$ to $\sim 10 M_\odot$; PBH fractions from $\sim 10^{-2}$ to $\sim 10^2$; peak GW abundances from $\sim 10^{-31}$ to $\sim 10^{-8}$; and peak GW frequencies from $\sim\unit[10^{-10}]{Hz}$ to $\sim\unit[10^{-2}]{Hz}$. Imposing the criterion that we only include the FOPT scenarios that lead to Doppler phase shifts within reach by SKA (labelled as ``SKA-PTA"), we find that the coverage of PBH masses and PBH fractions are significantly reduced as expected, since PTA is sensitive to PBH masses of at least $\sim 10^{-10}M_\odot$ and PBH fractions of at least $\sim 0.3$. On the other hand, recall from Eq. (\ref{aveMPBHest}) that the average PBH mass goes as the inverse square of the percolation temperature and is weakly dependent on the FOPT strength $\alpha_{tr}$; at the same time, a detectable signal in PTA and GW observations will require larger FOPT strengths at low percolation temperatures. Hence, the upper limit on the average PBH mass is a result of imposing a maximum $\alpha_{tr}$. As for GW observables, the $f_s$ histogram peaks at $\sim \unit[10^{-8}]{Hz}$, and covers a frequency band of $\sim 10^{-9}$ to $\sim \unit[10^{-7}]{Hz}$, corresponding to the effective frequency range of SKA. Focusing on the lower left corner of Fig. \ref{fig:HistPhys} where we have the histogram plot for $\Omega_s^{peak} h^2$, we see that we can still generate a wide range of GW abundances even after selecting those FOPT scenarios that produce both PTA signals and PBH fractions below unity; these FOPT scenarios are below the reach of GW searches in SKA. Further applying the filter where we take those points that also give a detectable GW signal in SKA, this narrows down the range of the values of the physical observables; in particular, average PBH masses within $\sim 10^{-8}$ to $10^{-5} M_\odot$ can generate signals in PTA and GW observations using SKA.

In Fig. \ref{fig:VeffScan} we present the results of our parameter scan as corner plots, where we conveniently show two dimensional slices of the entire scanned hypervolume. The red curves roughly enclose the points which correspond to FOPT scenarios where the resulting PBHs can generate a detectable Doppler shift in the pulsar timing signal from an SKA-like PTA observation. Those points that also generate GW signals within the sensitivity reach of SKA are bordered by the blue curves, and the subset of points where the PBH fraction is below unity are bounded by the green curves. The points outside the red curves correspond to either: (i) unphysical points which do not yield values for $T_0$ and/or $T_c$, according to Eq. (\ref{PhysicalCondition}), and hence do not lead to a FOPT; or (ii) points which are physical, but are beyond the reach of both pulsar timing probes of PBHs nor searches for stochastic GWs in SKA. We focused solely on filtering our scanned data points based on the sensitivity reach of SKA to both Doppler signals from transiting PBHs and GW signals from FOPT to highlight the potential of SKA to probe FOPT in the early Universe through these two observational signatures. Based on our scan, we can identify the class of effective potentials, which could come from some fundamental physics framework, that yield interesting phenomenological signatures which can be probed through SKA. Analogously, to provide us with insights on how the fundamental parameters affect the physical observables, we show a similar set of corner plots in Fig. \ref{fig:PhysCorrScan}. 

We can identify a few interesting features from the corner plots of Figs. \ref{fig:VeffScan} and \ref{fig:PhysCorrScan}. In Fig. \ref{fig:VeffScan}, we observe a lower limit for $B^{1/4}$ at $\sim \unit[1]{keV}$, which arises as a result of the physical conditions imposed in Eq. (\ref{PhysicalCondition}); similarly the white region in the range $10^{-1} \leq D \leq 1$ in the $\lambda$-$D$ panel can be attributed to an unphysical region where no $T_c$ can be obtained. As for the PBH masses, PBH fractions, and GW frequency, we see in Fig. \ref{fig:PhysCorrScan} that the main parameter that sets the scale for these physical parameters is $B^{1/4}$; note that we have traded $B^{1/4}$ with the destabilization temperature $T_0$, so that $B^{1/4}$ sets the temperature scale at which the FOPT occurs. Furthermore, we observe that the constraint on the PBH fraction below unity, for points that can be probed by SKA through pulsar timing and GW observations, restricts $B^{1/4}$, up to $\sim \unit[20]{keV}$. This can be understood by looking at the upper left corner of Fig. \ref{fig:PhysCorrScan}, where the PBH fraction appears to be positively correlated with $B^{1/4}$. Intuitively, this makes sense since a larger $B^{1/4}$ leads to a larger percolation temperature, which will increase the total energy density of the PBHs, leading to larger PBH fractions. Still in the same panel, the average PBH mass is anticorrelated with $B^{1/4}$; here, the average PBH mass is expected to decrease for higher percolation temperatures according to Eq. (\ref{aveMPBHest}). Hence, pushing for larger values of $B^{1/4}$ will generally bring us to lighter PBHs beyond the reach of PTA. 
\section{Conclusions}
\label{sec:conclusion}
In this paper, we studied the PBHs produced from the collapse of Fermi balls, which originate from DM particles filtered away from the walls of expanding true vacuum bubbles during a dark FOPT. Earlier work has shown that the generated PBHs follow a mass distribution, dictated by the distribution of the radii of false vacuum bubbles. In turn, the distribution is determined by the: percolation temperature; strength of FOPT; rate of FOPT; bubble wall velocity, assumed to be Chapman-Jouguet; and the DM asymmetry. We determined the sensitivity reach of pulsar timing arrays to the Doppler phase shift produced by these transiting PBHs, taking the case of monochromatic PBHs as a reference scenario and taking the average PBH mass to be at least $10^{-10} M_\odot$. Viable parameter points provide values of the SNR, a functional of the phase shift, above a threshold value set by the number of pulsars in the PTA. Taking the PTA $f_{PBH}-M_{PBH}$ sensitivity curve in the case of monochromatic PBHs, and identifying the monochromatic PBH mass with average PBH mass $\langle M\rangle$ from the PBH mass distribution, we were able to identify the corresponding FOPT parameters that are within reach by an SKA-like PTA measurement. As a cross check, we performed a Monte Carlo simulation of PBH masses sampled from the PBH mass distribution set by the viable FOPT parameters, and found that all of these parameter points lead to a situation where 90\% of the mock simulations produce a Doppler signal with an SNR above threshold. This can be explained by the fact that the mass distribution is sharply peaked, and thus most PBHs in the simulation have masses that hover around $\langle M\rangle$. Fixing $\eta_\chi = 10^{-5}$ and the ratio of the percolation temperature to the critical temperature to be 0.9, the rate of the FOPT ranges from $\beta/H_* \simeq 10$ to $10^4$, covering a wide range of values for the FOPT strength $\alpha_{tr}$, from $\sim 10^{-6}$ to $10^{-1}$, within the \unit[1-10]{keV} temperature range. Since the PBH fraction goes as $\Delta V_{eff}^{1/4}$, and since the PTA sensitivity curve requires a fairly constant PBH fraction for PBH masses above $10^{-8}M_\odot$, a lower percolation temperature implies a lower radiation energy density, requiring a larger FOPT strength. The case of $\eta_\chi = 10^{-4}$ and percolation temperatures 0.1-1 keV leads to similar sensitivity reach in $\alpha_{tr}$ and an order of magnitude increase in $\beta/H_*$. For $\eta_\chi = 10^{-5}$, and setting the percolation temperature to be fixed around the 1-10 keV range, we require at least $\beta/H_* \sim 10^3$ to produce average PBH masses below $10^{-8}M_\odot$. Below this PBH mass, PTA limits on the PBH fraction become weaker, which translates to larger values of the FOPT strength.

The same FOPT that produced the PBHs can also produce GW signals that can be probed through the correlated shifts of a collection of stars via pulsar timing. We concentrated our attention to GWs produced by sound waves since these provide the dominant contribution to the GW signal from FOPT. Their spectral shape is known and is determined by the peak GW abundance and peak frequency. We then obtained the peak-integrated sensitivity curves for THEIA, SKA, and $\mu$Ares, \textit{i.e.} the peak GW abundance as a function of the corresponding peak frequency corresponding to a GW signal with SNR = 1 and 20-year observation time. We found that the corresponding peak frequency of the GW signal falls within the sub-nHz to sub-$\mu$Hz range, and that THEIA offers the best sensitivity, while SKA offers sensitivity to our FOPT scenario that is comparable with THEIA. For projected SKA sensitivities to GWs, relevant FOPT rates lie in the range $10^2 \lesssim \beta/H_* \lesssim 10^4$ provides SNR $> 1$ for the benchmark cases $\eta_\chi = 10^{-5}$, keV-scale percolation temperatures, and $T_*/T_c = 0.9$. 

Finally we considered a class of generic effective quartic potentials that can realize the FOPT, from which we can directly calculate the FOPT parameters relevant for PBH formation and GW production. Performing a parameter scan over this class of effective potentials, fixing $\eta_\chi = 10^{-5}$, we identified points that lead to PBH Doppler signals within reach by an SKA-like PTA measurement, and also those that lead to GW sound wave production that can be seen by SKA. We found that there is a significant region of the parameter space that provides a detectable Doppler phase shift signal in an SKA-like PTA measurement, where stochastic GW searches in SKA may also cover. This may provide further motivation to utilize the distortions in the pulsar timing signal, beyond the search for stochastic GWs, to search for other signatures of FOPT in the early Universe, most notably the PBH formation discussed in this work. It is hoped that shifts in the timing signal in PTAs will be used not just to search for stochastic GWs, but also substructures that could be present in the Galactic neighborhood. The results of our scans can also be useful for constraining explicit models with an additional scalar field that can be mapped to our generic quartic potential.

\section*{Acknowledgments}
We acknowledge the kind support of the National Science and Technology Council of the Republic of China (formerly the Ministry of Science and Technology), with grant number NSTC 111-2811-M-007-018-MY2. This work used high-performance computing facilities operated by the Center for Informatics and Computation in Astronomy (CICA) at National Tsing Hua University. This equipment was funded by the Ministry of Education of Taiwan, the National Science and Technology Council of Taiwan, and National Tsing Hua University. The authors would like to thank Reginald Bernardo, Thong Tran Quang Nguyen, Martin Spinrath, and Yu-Min Yeh for the helpful discussions and comments. 
%%%%%%%%%%%%%%%%%%%%%%%%%%%%%%%%%%%%%%%%%%%%%%%%%%%%%%%%%%%%%%%%%%%%%%%%%%%%%%%%%%%%%%%%%%%%%%%%%%%%%%%%%%%%%%%%%%%%%%%%%%%%%%%%%%%%%%%%%%%%%%%%%%%%%%%%%%%%%%%%%%%%%%%%%%%%%%%%%%%%
\appendix
\section{Doppler and Shapiro shifts in frequency and phase for pointlike PBH}
\label{appendix:derivations}
We begin by deriving explicit expressions for the fractional change in frequency for the Doppler and Shapiro signals, in the case of transiting monochromatic PBHs. The Doppler and Shapiro shifts are given by \cite{Dror:2019twh}
\begin{eqnarray}
\label{dopplermain}\left(\frac{\delta \nu}{\nu}\right)_D &=& \frac{1}{c}\int dt~\hat{d}\cdot\vec{\nabla}\Phi,\\
\label{shapiromain}\left(\frac{\delta \nu}{\nu}\right)_S &=& -\frac{2}{c^3}\int_{LOS} dz~\vec{v}\cdot\vec{\nabla}\Phi,
\end{eqnarray}
where $\hat{d}$ is the unit vector pointing from the Earth to the pulsar, and $\Phi$ is the gravitational potential due to the passing DM substructure or PBH. For the Doppler shift, the gradient of the potential is evaluated at the location of the pulsar, while, for the Shapiro shift, it is evaluated at a point along the LOS. Note that in both cases, we assume that the velocity $\vec{v}$ of the passing object is constant.\\
\newline
\noindent For the Doppler shift, we denote $\vec{r}$ to be the position vector pointing from the pulsar to the transiting object. We can write it as
\begin{eqnarray}
\label{trajectory}\vec{r} = \vec{b} + \vec{v}(t-t_{D}),
\end{eqnarray} 
where $\vec{b}$ is the vector pointing to the closest approach along the object's trajectory, and $t_D$ is the time of closest approach. The gravitational potential at the pulsar's location is
\begin{eqnarray}
\Phi = -\frac{GM_{PBH}}{r},
\end{eqnarray}
so that
\begin{eqnarray}
\vec{\nabla}\Phi = \frac{GM_{PBH}}{r^2}\hat{r}.
\end{eqnarray}
Using (\ref{dopplermain}), we have
\begin{eqnarray}
\left(\frac{\delta \nu}{\nu}\right)_D = \frac{GM_{PBH}}{c}\int dt~\frac{\hat{d}\cdot\hat{r}}{r^2}.
\end{eqnarray}
From the trajectory (\ref{trajectory}), we have
\begin{eqnarray}
r^2 &=& b^2 + v^2(t-t_D)^2 \equiv b^2 (1+x_D^2)\\
\hat{r} &=& \frac{\hat{b}}{\sqrt{1+x_D^2}}+\hat{v}\frac{x_D}{\sqrt{1+x_D^2}}.
\end{eqnarray}
Here we have introduced the variable
\begin{eqnarray}
x_D \equiv \frac{v(t-t_D)}{b}.
\end{eqnarray}
Then we have
\begin{eqnarray}
\left(\frac{\delta \nu}{\nu}\right)_D = \frac{GM_{PBH}}{cb^2}\frac{b}{v}\int dx_D~\left[(\hat{d}\cdot\hat{b})\frac{1}{(1+x_D^2)^{3/2}}+(\hat{d}\cdot\hat{v})\frac{x_D}{(1+x_D^2)^{3/2}}\right].
\end{eqnarray}
Integrating the above expression, and ignoring the overall constant, we have
\begin{eqnarray}
\left(\frac{\delta \nu}{\nu}\right)_D = \frac{GM_{PBH}}{v^2 \tau_D c}\left[(\hat{d}\cdot\hat{b})\frac{x_D}{(1+x_D^2)^{1/2}}-(\hat{d}\cdot\hat{v})\frac{1}{(1+x_D^2)^{1/2}}\right],~\tau_D \equiv \frac{b}{v}.
\end{eqnarray}
The phase is the integral of the frequency shift over time, and is just
\begin{eqnarray}
\delta\phi_D(t) = \frac{GM_{PBH} \nu}{v^2 c}\left[\left(\hat{d}\cdot\hat{b}\right)\sqrt{1+x_D^2}-\left(\hat{d}\cdot\hat{v}\right)\sinh^{-1} x_D\right].
\end{eqnarray}
In terms of the initial position $\vec{r}_0$, we have
\begin{eqnarray}
t_D = -\frac{\vec{r}_0 \cdot \vec{v}}{v^2}&,&~\tau_D = \frac{\sqrt{r_0^2 - v^2 t_D^2}}{v},\\
\hat{d}\cdot \hat{b} &=& \left(\hat{d}\cdot \hat{r}_0\right)\frac{r_0}{v\tau_D} + \left(\hat{d}\cdot \hat{v}\right)\frac{t_D}{\tau_D}.
\end{eqnarray}
As for the Shapiro signal, it is convenient to work in cylindrical coordinates, so that the position of the transiting object can be written as
\begin{eqnarray}
\vec{r} = s(t)\hat{s}+z(t)\hat{z},
\end{eqnarray}
where $0 < z < L$, and $L$ is the distance between the Earth and the pulsar. The potential at a point $z'$ along the LOS ($z'=0$ is taken to be the pulsar's location) is
\begin{eqnarray}
\Phi = -\frac{GM_{PBH}}{\sqrt{s^2+(z'-z)^2}},
\end{eqnarray}
so that
\begin{eqnarray}
\vec{\nabla}\Phi = \frac{GM_{PBH}}{[s^2+(z'-z)^2]^{3/2}}\left[s\hat{s}-(z'-z)\hat{z}\right].
\end{eqnarray}
From (\ref{shapiromain}) we have
\begin{eqnarray}
\nonumber\left(\frac{\delta \nu}{\nu}\right)_S = \frac{2GM_{PBH}}{c^3}\int_0^{L} dz'\left\{(\vec{v}\cdot\hat{s})\frac{s}{[s^2+(z'-z)^2]^{3/2}}-(\vec{v}\cdot\hat{z})\frac{(z'-z)}{[s^2+(z'-z)^2]^{3/2}}\right\}.\\
\end{eqnarray}
Performing the integral yields
\begin{eqnarray}
\nonumber\left(\frac{\delta \nu}{\nu}\right)_S = \frac{2GM_{PBH}}{c^3}\left\{(\vec{v}\cdot\hat{s})\frac{(z'-z)}{s[s^2+(z'-z)^2]^{1/2}}+(\vec{v}\cdot\hat{z})\frac{1}{[s^2+(z'-z)^2]^{1/2}}\right\}\Bigg\vert_{z'=0}^{z'=L}.\\
\end{eqnarray}
We then introduce the following approximation: since the Shapiro signal is relevant when the transiting object is sufficiently close to the LOS, we take $s \ll z, L, L-z$. Then we have
\begin{eqnarray}
\frac{(z'-z)}{s[s^2+(z'-z)^2]^{1/2}} \approx \frac{1}{s}\text{sgn}(z'-z),~\frac{1}{[s^2+(z'-z)^2]^{1/2}} \approx \frac{1}{\vert z'-z\vert}.
\end{eqnarray}
Because of the above expressions, we can drop the term proportional to $\vec{v}\cdot \hat{z}$, so that we have
\begin{eqnarray}
\left(\frac{\delta \nu}{\nu}\right)_S \approx \frac{4GM_{PBH}}{c^3}\frac{\vec{v}\cdot\hat{s}}{s}.
\end{eqnarray}
Let us then rewrite the velocity vector as
\begin{eqnarray}
\vec{v} = v_z \hat{z} + \vec{v}_\perp,
\end{eqnarray}
where $\vec{v}_\perp$ is the velocity component that lies on the plane, containing the transiting object, that is perpendicular to the LOS. By analogy with what we have done in writing the position vector for Doppler, the vector $\vec{s}$, which points from the LOS to the transiting object, can be written as
\begin{eqnarray}
\vec{s} = \vec{b}_\perp + \vec{v}_\perp (t-t_S),
\end{eqnarray}
so that
\begin{eqnarray}
s = b_\perp\sqrt{1+x_S^2},~x_S \equiv \frac{v_\perp (t-t_S)}{b_\perp},\\
\hat{s} = \frac{\hat{b}_\perp}{\sqrt{1+x_S^2}}+\hat{v}_\perp \frac{x_S}{\sqrt{1+x_S^2}},
\end{eqnarray} 
and thus
\begin{eqnarray}
\left(\frac{\delta \nu}{\nu}\right)_S \approx \frac{4GM_{PBH}}{\tau_S c^3}\frac{x_S}{(1+x_S^2)},~\tau_S \equiv \frac{b_\perp}{v_\perp};
\end{eqnarray}
the associated phase shift is just
\begin{eqnarray}
\delta\phi_S(t) = \frac{2GM_{PBH} \nu}{c^3}\ln\left(1+x_S^2\right).
\end{eqnarray}
Analogous to the Doppler case, if we are given the initial cylindrical radial position $\vec{s}_0$, we have
\begin{eqnarray}
\vec{v}_\perp = \vec{v} - \hat{d}(\hat{d}\cdot\vec{v}) &,& v_\perp = \vert \vec{v}\times \hat{d}\vert\\
t_S = -\frac{\vec{v}_\perp\cdot\vec{s}_0}{v_\perp^2}&,&~\tau_S = \frac{\sqrt{s_0^2-v_\perp^2 t_S^2}}{v_\perp}
\end{eqnarray}
In all of the above expressions, we assume a PBH point source. Extending this to diffuse objects is just a matter of introducing a form factor to the signals; in \cite{Ramani:2020hdo} the frequency shift is multiplied by
\begin{eqnarray}
\label{ffac}\mathcal{F}(M,x) = x \int_0^\infty dk~W(k,M) J_1(kx),
\end{eqnarray}
evaluated at $x = b$ for Doppler or $x = b_\perp$ for Shapiro. Note that
\begin{eqnarray}
W(k,M) \equiv \frac{1}{M}\int d^3\Vec{r}~ e^{i\Vec{k}\cdot \Vec{r}}\rho(r),
\end{eqnarray}
assuming that the density profile of the diffuse object is spherically symmetric.

% Bibliography

%% [A] Recommended: using JHEP.bst file
 \bibliographystyle{JHEP}
 \bibliography{JHEP_JTA_PYT_ManuscriptRevised.bib}

%% or
%% [B] Manual formatting (see below)
%% (i) We suggest to always provide author, title and journal data or doi:
%% in short all the informations that clearly identify a document.
%% (ii) please avoid comments such as "For a review'', "For some examples",
%% "and references therein" or move them in the text. In general, please leave only references in the bibliography and move all
%% accessory text in footnotes.
%% (iii) Also, please have only one work for each \bibitem.

\end{document}